\documentclass[structabstract]{aa} 

\usepackage{graphicx}
\usepackage[varg]{txfonts}

\usepackage{natbib}
\usepackage{multirow}

\begin{document} 

	\title{Indication for an intermediate-mass black hole in the globular cluster NGC 5286  from kinematics   
                                                     }
	
	
	\author{A. Feldmeier
			\inst{\ref{inst1}}
		\and N. L{\"u}tzgendorf
			\inst{\ref{inst1}}
		\and N. Neumayer
			\inst{\ref{inst1}}
		\and M. Kissler-Patig
			\inst{\ref{inst2}}$^,$\inst{\ref{inst1}}
		\and K. Gebhardt
			\inst{\ref{inst4}}
			\and H. Baumgardt
			\inst{\ref{inst3}}
		\and E. Noyola
			\inst{\ref{inst6}}
		\and P. T. de Zeeuw
			\inst{\ref{inst1}}$^,$\inst{\ref{inst7}}
		\and B. Jalali
			\inst{\ref{inst5}}
				}

 \institute{European Southern Observatory (ESO), Karl-Schwarzschild-Stra{\ss}e 2, 85748 Garching, Germany   
 	\\
	\email{afeldmei@eso.org}
		\label{inst1}
          \and 
          Gemini Observatory, 670 N. A'ohoku Place, Hilo, Hawaii, 96720, USA 
          	\label{inst2}
                  \and 
          Astronomy Department, University of Texas at Austin, Austin, TX 78712, USA 
          	\label{inst4}
	  \and 
          School of Mathematics and Physics, University of Queensland, Brisbane, QLD 4072, Australia 
          	\label{inst3}
                 \and 
          Instituto de Astronomia, Universidad Nacional Autonoma de Mexico (UNAM), A.P. 70-264, 04510 Mexico 
          	\label{inst6}
          \and 
          Sterrewacht Leiden, Leiden University, Postbus 9513, 2300 RA Leiden, The Netherlands 
          	\label{inst7}
	   \and 
          I. Physikalisches Institut, Universit{\"a}t zu K{\"o}ln, Z{\"u}lpicher Str. 77, 50937 K{\"o}ln, Germany 
          	\label{inst5}
          }

\date{Received xxxxxxxx xx, xxxx; accepted xxxxxxx xx, xxxx}

 \abstract
   {Intermediate-mass black holes (IMBHs)   fill the gap between stellar-mass black holes and supermassive black holes (SMBHs). The existence of the latter is widely accepted, but there are only few detections of intermediate-mass black  holes (10$^{2}$ - 10$^{5}$ M$_{\odot}$) so far.  
 Simulations have shown that intermediate-mass black holes may form in dense  star clusters, and therefore may still be present in these smaller stellar systems. Also, extrapolating the $M_{\bullet}$ - $\sigma$ scaling relation to lower masses predicts intermediate-mass black holes in systems with $\sigma$ $\sim$ 10 - 20 km/s such as globular clusters. }
   {We investigate the Galactic globular cluster NGC 5286 for indications of a central intermediate-mass black hole using spectroscopic data from VLT/FLAMES\thanks{Based on observations collected at the European Organisation for Astronomical Research in the Southern 	Hemisphere, Chile \mbox{(085.D-0928})}, velocity measurements from the Rutgers Fabry Perot (RFP) at CTIO, and photometric data from HST/ACS.}
   {We compute the photometric center, a surface brightness  profile, and a velocity-dispersion profile. We run analytic spherical and axisymmetric Jeans models with different central black-hole masses, anisotropy, mass-to-light ratio, and inclination.  Further, we  compare the data to a grid of $N$-body  simulations without tidal field. Additionally, we use one $N$-body  simulation to check the results of the spherical Jeans models for the total cluster mass.}
   {Both the Jeans models and the $N$-body  simulations favor  the presence of  a central black hole in NGC 5286  and our detection is at the 1- to 1.5-$\sigma$ level. From the spherical Jeans models we obtain  a best fit with black-hole mass M$_\bullet$ = (1.5 $\pm$1.0)\,$\cdot$\,10$^3$\,M$_\odot$. The error is the 68\% confidence limit from Monte Carlo simulations. Axisymmetric models give a consistent result.  The best fitting $N$-body  model is found with a black hole of 0.9\% of the total cluster mass (4.38$\,\pm\,$0.18)\,$\cdot$\,10$^5$\,M$_{\odot}$, which results in an IMBH mass of   M$_\bullet$\,=\,(3.9$\,\pm\,$2.0)\,$\cdot$\,10$^3$\,M$_\odot$.  Jeans models give values for the  total cluster mass that are lower by up to 34\% due to a lower value of M/L. Our test of the Jeans models with $N$-body simulation data shows that  the discrepancy in the total cluster mass has two reasons:  The influence of a radially varying M/L profile, and underestimation of the velocity dispersion  as  the measurements are limited to bright stars, which have lower velocities than fainter stars. We conclude that detection of IMBHs in Galactic globular clusters remains a challenging task unless their mass fractions are above those found for SMBHs in nearby galaxies.}
{}

   \keywords{Black hole physics --
                globular clusters: individual: (NGC 5286) --
                stars: kinematics and dynamics
               }

  \maketitle

\section{Introduction}

The existence of supermassive black holes (SMBHs) is  widely accepted. Mass estimates come either from kinematic measurements, or from X-ray and radio luminosities, which assume certain accretion rates. However, the formation and growth of SMBHs is not well understood, although it is somehow linked to the evolution of the environment. 
The mass of a central  black hole is correlated to the mass of the bulge $M_{b}$ of its host system (e.g.  \citealt{nadine}), and to the stellar velocity dispersion $\sigma$ of the bulge (e.g. \citealt{msigma,ferrarese,gueltekin}). Systems with higher $M_{b}$ and higher $\sigma$ contain more massive central black holes.  This suggests a connection between  the  formation and evolution  of central black holes and  their host systems.

There are several detections of quasars at a redshift higher than 6  (e.g. \citealt{fan,willott,quasar7}). The  masses for the central black holes as inferred from the luminosity are of the order of $ \sim$\,10$^9\,$M$_\odot$. These high masses  cannot be explained by accretion onto a stellar-mass black hole, as not enough time has elapsed at $z\sim$ 6: the accretion rates would greatly exceed  the Eddington limit.  In order to build a black hole of 10$^9\,$M$_\odot$ at $z\sim$ 6, the black-hole seed  requires a mass of 10$^2 $-$ 10^5$\,M$_\odot$, as shown by \cite{volonteri}. Black holes in this mass range are called intermediate-mass black holes (IMBHs) and must have formed in the young universe. 

There are theories for the formation of IMBHs from Population III stars \citep{madau, bromm}, gas-dynamical processes \citep{loeb}, and from stellar-dynamical instabilities in star clusters \citep{imbhmerge}. The latter theory is of particular interest for globular clusters. In very compact star clusters the most massive stars sink to the core due to mass segregation. If the cluster is very dense,  this occurs before the stars explode and evolve to  compact remnants ($\sim 10^{6}$\,yr, \citealt{imbhmerge}).  In a dense core  the massive stars undergo  runaway stellar mergers.  After  several collisions, a very massive star (VMS) with \textgreater\,100\,M$_\odot$ forms and grows quickly (\citealt{zwart}). 
\cite{yungelson} used an evolutionary code to compute  the mass loss for VMSs of initial solar metallicity ($Z$ = 0.02). They found that the mass loss due to stellar winds is so high that it is unlikely that the VMS  forms an IMBH.   \cite{belkus} studied the evolution of VMSs at different metallicities, and found that low-metallicity stars with $Z$ = 0.001 suffer less mass loss. Their final mass is by a factor of $\sim$ 2$-$3 larger than for stars with a metallicity of $Z$ = 0.02. These VMSs can evolve to IMBHs with  some 10$^{2}$\,M$_{\odot}$.  The central black hole continues accreting nearby stars, until the reservoir of massive stars is exhausted.  
 
 If a cluster is not dense enough,  stellar evolution is faster than  mass segregation. Massive stars evolve to stellar remnants before reaching the center. 
\cite{miller} proposed that successive  mergers of  stellar remnants and accretion can form an IMBH in a globular cluster. 
Recent models examined the retention fraction of stellar black holes in star clusters and merger rates due to gravitational radiation and found different results. The retention fraction for black holes binaries is only 1\% to 5\% according to \cite{moody}.  \cite{millerdavies}  found that nearly all black holes are kicked out during mergers. On the other hand, \cite{morscher} came to the result that less than 50\% of the  black holes are dynamically ejected, and the total merger rate is $\sim$ 1 per Gyr and globular cluster. Also \cite{leary} found significant probability of black hole growth. However, runaway mergers of black holes to form an IMBH seem to be less likely for globular cluster-sized systems.

Globular clusters orbiting around galaxies will sink to the galactic center due to dynamical friction \citep{imbhmerge}. While stars in the outskirts of the clusters are stripped away from the galaxy's tidal field and settle in the galactic bulge, the central IMBHs can be carried to the center of the galaxy. They can form binary IMBHs. Some IMBHs are ejected in three-body interactions, and some IMBHs   merge through gravitational radiation into more massive black holes, and form a SMBH by successive mergers (\citealt{imbhmerge}). If  the $M_{\bullet}$ - $\sigma$ scaling relation holds  down to the low-mass end, globular clusters, which have velocity dispersions of $\sim$\,10 - 20\,km/s, are a possible environment for intermediate-mass black holes.

\cite{NG08} found kinematic signatures for a black hole of (4.0 $^{+0.75}_{-1.0}$)\,$\cdot$\,10$^4$\,M$_\odot$ in $\omega$\,Centauri from the radial velocity dispersion and isotropic Jeans models. Their result was challenged by \cite{vdM10}, who used HST proper motion measurements and determined an upper limit of 1.8\,$\cdot$\,10$^4$\,M$_\odot$. The groups used different center coordinates, which is probably one of the main reasons for this discrepancy. With the kinematic center from  \cite{NG10},  \cite{behrang} performed direct $N$-body  simulations  and found a central IMBH with $\sim$ 5\,$\cdot$\,10$^4$\,M$_\odot$, which is consistent with the spherical isotropic models of  \cite{NG10}. 
Another good candidate with kinematic measurements is NGC 6388.  \cite{Nora} used integral field unit data and Jeans models to determine a black-hole mass of (1.7\,$\pm$\,0.9)\,$\cdot$\,10$^4$\,M$_\odot$. The most massive globular cluster in M31, G1, shows kinematic signatures \citep{G1} as well as  X-ray  \citep{g1xray} and radio  emission \citep{g1radio}, all consistent with an IMBH of $\sim 2 \cdot 10^4$\,M$_\odot$.  \cite{millerg1} confirmed the detection of X-ray emission but detected no radio  emission in G1. 
They concluded that the detected emission comes more likely from a low-mass  X-ray binary (LMXB) with a stellar-mass black hole than from an IMBH.

Recent studies  found no significant radio sources in the center of $\omega$\,Centauri \citep{omradio} and NGC 6388 \citep{6388radio}. However, mass estimates from upper limits on radio emission require assumptions on the gas properties and accretion model.  
\cite{pepe} showed that  the accretion rate in globular clusters depends  on  the mass of the cluster and its velocity dispersion. Further, \cite{perna} found that the widely used Bondi-Hoyle accretion rate \citep{bondi} is an upper limit of  the true mass accretion, which can be lower by several orders of magnitude. The fraction of the Bondi accretion rate at which accretion occurs is highly uncertain \citep{maccarone}.

There are further examples for black holes in extragalactic globular clusters.    \cite{zepf} measured broad emission lines  in the globular cluster RZ 2109 in the Virgo elliptical galaxy NGC 4472 and conclude that, in combination with the measured X-ray luminosity, this can be explained by the presence of a  $\sim$ 10\,M$_\odot$ black hole. A globular cluster in  NGC 1399 shows broad emission lines and X-ray emission  that can be modeled with the  tidal disruption of a horizontal branch star by a massive (50$-$100\,M$_\odot$) black hole \citep{clausen}.

\cite{HBGC} performed $N$-body simulations and ascertain that globular clusters with an IMBH have a rather large core, and a shallow cusp in the surface brightness profile.  The  velocity-dispersion profile rises towards the center, but this is difficult to detect if only the brightest stars of a cluster are observed. 
\cite{NG06} obtained a surface brightness profile of NGC 5286 that shows a shallow cusp, and for this reason we set out to search for kinematic evidence for an IMBH.

NGC 5286 is in the halo of the Milky Way at a distance of about 11.7\,kpc  (\citealp{harris}).  Table\,\ref{tab:5286data} lists some properties of this cluster.
\cite{binacs} found some foreground stars   in the region of NGC 5286. The fairly bright (m$_V$ = 4.6) spectroscopic binary star M~Centauri (HR 5172)  is at a distance of about 4\arcmin\, from the center of NGC 5286. \cite{cmd09}  compared the CMD of NGC 5286 with M3 and found  that NGC 5286 is about (1.7\,$\pm$\,0.9)\,Gyr older than M3. It may belong to the oldest globular clusters in our Galaxy. \cite{crane} argued that NGC 5286 is associated with  a ringlike structure near the Galactic anticenter, based on its position and kinematics. 
This ringlike filament of stars around the Milky Way,  the so-called Monoceros ring,  can be the tidal stream of a disrupted dwarf galaxy, with the nucleus  in Canis Major (\citealt{canismajor}). But the existence of the Canis Major dwarf galaxy is still under debate. However, it is possible that NGC 5286 is originally a globular cluster of the Canis Major dwarf galaxy and not of the Milky Way.
 \begin{table}
 \caption{Properties of the globular cluster NGC 5286. References: Go10~=~\cite{GoldCen}, Ha10~=~\citet[2010 edition]{harris}, Zi80~=~\cite{reddening}, Zo10~=~\cite{varstar}, Wh87~=~\cite{white}}
  \label{tab:5286data}
\centering
\begin{tabular}{l l l}
\noalign{\smallskip}
\hline\hline
\noalign{\smallskip}
Parameter&Value&Reference \\
\noalign{\smallskip}
\hline
\noalign{\smallskip}
 RA $\alpha$ (J2000)&13h 46m 26.81s&Go10\\
 DEC $\delta$ (J2000)&-51\degr\,22\arcmin\,27.3\arcsec&Go10\\
 Galactic longitude l&311.6142\degr&Go10 \\
 Galactic latitude b& +10.5678\degr&Go10 \\
 Distance from the sun& 11.7  kpc	&Ha10 \\
 Distance from the Galactic center&8.9 kpc& Ha10  \\
 Foreground reddening E(B$-$V)&0.24	&Zi80  \\
 Absolute visual magnitude M$_V$& -8.74 & Ha10  \\
 Metallicity [Fe/H]&-1.67 &Zo10 \\ 
 Projected ellipticity  $\epsilon$\,=\,1$-$(b/a)&0.12&Wh87 \\
 Heliocentric radial velocity v$_{r}$&(57.4 $\pm$ 1.5) km/s& Ha10  \\
Central velocity dispersion $\sigma_{c}$	& (8.1 $\pm$ 1.0) km/s&Ha10 \\
Concentration c\,=\,$\log(r_{t}/r_{c})$& 1.41& Ha10\\
Core radius r$_{c}$&16.8\arcsec&Ha10 \\
Half-light radius r$_{h}$& 43.8\arcsec&Ha10 \\
\noalign{\smallskip}
 \hline
\noalign{\smallskip}
\end{tabular}
\end{table}

This paper is organized as follows: In Section \ref{sec:photo} we describe the photometric analysis, which includes the center determination and calculation of a surface brightness profile. This gives information on the stellar distribution and density. The observation and data reduction of the spectroscopic FLAMES data are outlined in Section \ref{sec:spec}. With these data we calculate a velocity map and a velocity-dispersion profile for the central 25\arcsec, as described in Section \ref{sec:kin}. For the outer part of the profile we use a data set obtained from the Rutgers Fabry Perot, and we investigate it for indications of rotation. Section \ref{sec:jeans} and Section \ref {sec:nb} introduce the results of the  Jeans models and $N$-body simulations, which are compared to the data.  
 We summarize and discuss our results in Section \ref{sec:end}.

\section{Photometry}
\label{sec:photo}
The photometric data for NGC 5286 was obtained from the Hubble Space Telescope (HST) archive. Observations were made with the Advanced Camera for Surveys (ACS) in the Wide Field Channel (WFC) with a pixel scale of 0.05\,\arcsec/pixel. NGC 5286 is  one of 65 observed clusters, being part of the ``ACS Survey of Globular Cluster'' project  (\citealt{SaraACS1}). The data were reduced by the ``ACS Survey of Globular Cluster''-team as described in \cite{SaraACS1} and \cite{AndersonACS5}, and they also produced a star catalog, which provides a nearly complete list of all stars in the central 2\arcmin\,of the cluster, with information on position, V- and I-band photometry, and some data quality parameters for each star. The list for NGC 5286 contains 210,729 stars and was retrieved from the ``ACS Survey of Galactic Globular Clusters" database, whereas the HST reference image  \emph{hlsp\_acsggct\_hst\_acs-wfc\_ngc5286\_f606w\_v2\_img.fits} in the V-band was downloaded from  the HST-Archive.

\subsection{Center determination}

\label{sec:adopt}
It is essential to have  precise center coordinates, as the center influences the shape of the surface brightness profile and of the velocity-dispersion profile. Using the wrong center usually results in a shallower inner surface brightness profile, and this influences the outcome of our models. 

	The center of NGC 5286 was already determined by \cite{NG06}  and \cite{GoldCen}. 	With HST/WFPC2 data from the F555W filter (V-band) \cite{NG06} minimize the standard deviation of  star counts in eight  segments of a circle. \cite{GoldCen} search for the most symmetric point by counting stars in circular segments and fit ellipses to isodensity contours, using the same star catalog as we do. The center coordinates in right ascension $\alpha$ and declination $\delta$ are shown in Table \ref{tab:centers} relative to our reference image. We apply three approaches for the center determination.
\begin{table}
\caption{Center coordinates  (J2000)}
 \label{tab:centers}
 \centering
\begin{tabular}{l l l l}
\noalign{\smallskip}
\hline\hline
\noalign{\smallskip}
Center determination &$\alpha$&$\delta$&Error\\
& (h:m:s)& (\degr\,:\,\arcmin\,:\,\arcsec)& (\arcsec)\\
\noalign{\smallskip}
\hline
\noalign{\smallskip}

\cite{NG06}&13:46:26.844&-51:22:28.55	&0.5\\

\cite{GoldCen}&13:46:26.831&-51:22:27.81& 0.1\\

  Star counts&13:46:26.834&-51:22:27.94&0.3\\

  Cumulative distribution&13:46:26.844  &   -51:22:27.88   &  0.4\\

  Isodensity contours&13:46:26.846  &   -51:22:27.98    & 0.4\\
 \noalign{\smallskip}
  \hline
\noalign{\smallskip}
\end{tabular}

\end{table}

The first method is similar to the so-called ``Pie-slice Contours" method described by \cite{GoldCen} and \cite{Nora}. It uses star counts in opposing wedges.  
In a second approach we create a cumulative radial distribution of the stars, as described by  \cite{centertuc} and \cite{Nora}. 
The crucial point of the center determination is to decide in which magnitude interval we count the stars, i.e. where we make the magnitude cut. This problem arises from the star catalog construction. In the vicinity of bright stars it is not possible to detect fainter stars,  and the star list is therefore incomplete.  The resulting  holes in the apparent stellar distribution become a problem as we count the stars, and opposing wedges seem to be asymmetric. It is therefore important to use only stars brighter than a certain magnitude. In addition, the magnitude cut must not be too low, as then there would not be enough stars for a statistically meaningful result and the sample sizes would no longer be useful. The uncertainty from the magnitude cut is about 10 times larger than  the  uncertainty from the number of wedges. 
Therefore, we  use a combination of the different parameters and weight the results with the respective errors. This is done for the simple star count method as well as for the cumulative radial distribution method. The number of wedges we use are between 4 and 16, and we determine the median and standard deviation $\sigma_{med}$ of the resulting  center coordinates. We apply this method to magnitude cuts of 19, 20, 21, 22, and 23, and weight the resulting center coordinate with the respective standard deviation $\sigma_{med}$ to determine a center, and calculate the standard deviation $\sigma_{cut}$ of the weighted center. As a result we obtain center coordinates from the star count and the cumulative radial distribution method and use $\sigma_{cut}$ as a measure for the uncertainties. 

Alternatively, we determine the center with an approach that  is based on the ``Density Contour'' method as described in \cite{GoldCen}.  The center is found by creating isodensity contours and fitting ellipses to the contours. At first we create a grid of 200\arcsec\,x 200\arcsec\,with a grid point every 2\arcsec, and we use the center determined by \cite{NG06} as our grid center.   Assuming the center determined in \cite{GoldCen} as the grid center changes the result by less than 0.04\arcsec. At every grid point we apply a circle with a radius of 500 pixel (25\arcsec), find the stars inside the circle, and determine the density inside the circle around the grid point. Large radii of the circle result in smooth contour plots, but if the radius is too large, it reaches the edge of our star field. A contour plot is created with eight contour levels, and four ellipses are fit to the contours. The innermost and the outer three contours were ignored as they can be either biased by shot noise, or  be incomplete. For the remaining four contours the center of the ellipse fit is determined, and the median is adopted as the new center, with  the standard deviation  of the ellipse centers as estimated uncertainty.
Again, the choice of a magnitude cut affects the  outcome of the center coordinates. Excluding faint stars  smoothens the distribution, as fainter stars are not likely to be found in the vicinity of bright stars, and create under-dense regions around bright stars. This incompleteness is larger in the center of the cluster, as there are more bright stars. The difference of the center position is up to 3\arcsec\,for magnitude cuts at 15 and 29. We  combine the isodensity method for five magnitude cuts from 19 to 23. The difference in the center determination for these magnitude cuts is less than 0.8\arcsec, and the result therefore more robust. We calculate an error-weighted average, which is adopted as the new center. The standard deviation is our estimated error, and the results are listed in Table \ref{tab:centers}. 

\begin{figure}
\resizebox{\hsize}{!}{\includegraphics{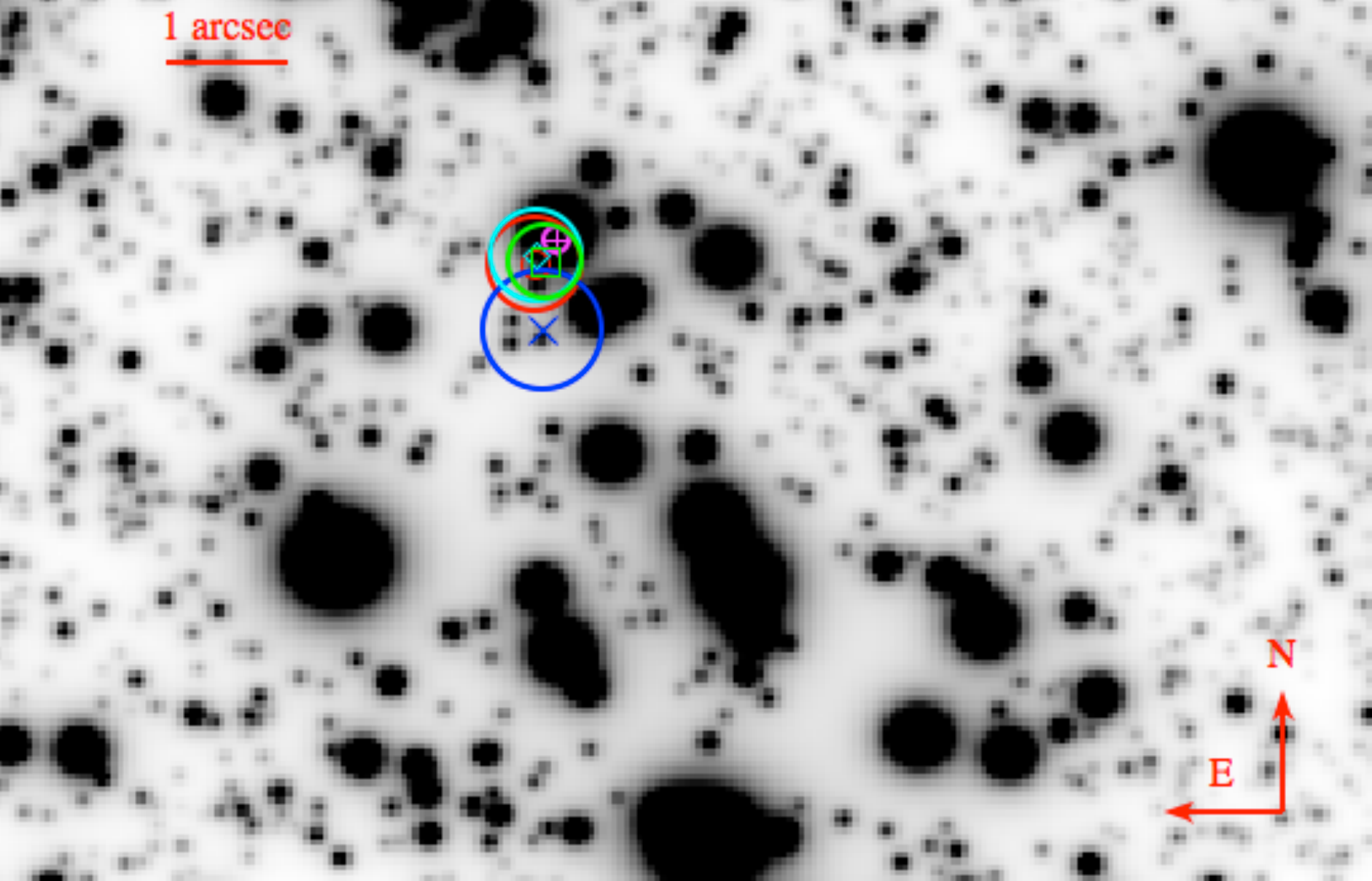} }
\caption{Finding chart for the different center coordinates with uncertainties. \cite{NG06}: cross, \cite{GoldCen}: plus sign, star count method: square, cumulative stellar distribution: diamond,  isodensity contours: circle point. The underlying image is a synthetic I-band image (see Section \ref{sec:velmap}).}
\label{fig:allcen}
\end{figure}
Figure \ref{fig:allcen} illustrates all centers from our  methods with respective errors plotted on top of a synthetic I-band ACS image (see Section \ref{sec:velmap}). They  all lie within their error ranges, and all of them overlap within their errors with  the uncertainty of the center derived by \cite{NG06}, and with the center of \cite{GoldCen}. This is not unexpected, as we use the same star catalog as \cite{GoldCen} and similar approaches to determine the center. 
As \cite{GoldCen} report, the center derived with the isodensity contour method is more reliable over a wide range of density distributions. The star count methods rely on the symmetry of the cluster, and depend more on sample size and cluster size, so we  adopt the center calculated with the isodensity contour method. By means of the  HST reference image   we find the final position in pixel,  right ascension $\alpha$, and declination $\delta$:
\begin{equation}
(x_{c}, 	\, y_{c})=(2934.80,     \,   2917.59)\, \pm\, (7.84, \,	5.88) \,\mathrm{pixel}
\end{equation}
\begin{equation}
\alpha =\,\,13\,: \, 46\,:\,26.847, \,\Delta\alpha=0.4\arcsec  \,\mathrm{(J2000)}
\end{equation}
\begin{equation}
 \delta = -51\,:\, 22\,:\, 27.954, \,\Delta\delta=0.3\arcsec
\end{equation}
As the distance of NGC 5286 is about 11.7\,kpc, the uncertainty of the center determination is only $\sim$\,0.022\,pc.					

\subsection{Surface brightness profile}
\label{sec:SB}
The next step of the photometric analysis is to calculate the surface brightness profile. In the literature there are already surface brightness profiles of NGC 5286, one of them is from \cite{NG06}, the other is from \cite{Trager}. \cite{NG06} use integrated light measurements, whereas our method is based on star counts in combination with  integrated light measurement as  described in \cite{Nora}. We use the HST reference image and star catalog in the V-band, and our center as determined in Section \ref{sec:adopt} via isodensity contours.
A combination of integrated light measurement and star counts makes sure that the surface brightness profile is complete and robust. Since the star catalog does not contain every faint star, using only star count measurements would neglect the light contribution from faint stars. But their contribution to the light is taken into account in integrated light measurements. To use only integrated light causes problems in the determination of the surface brightness at the center. Bright stars produce shot noise and the resulting profile is sensitive to the choice of the bins. This effect is weaker with star counts, and therefore we combine both methods, to make our profile more robust. 

For the star counts we apply circular bins around the center,  sum the fluxes of all stars with 14\textless m$_{V}$\textless19, and divide by the area of each bin. 
The integrated light measurement is applied for fainter stars with m$_{V}$$\geq$19, and the brighter stars are masked out in the reference image using a circular mask with a radius of 0.25\arcsec\,(5 HST pixels). 
We use the same circular bins for both methods and use the bi-weight to determine the distribution of pixel counts, and sum the fluxes of both methods.

To obtain  a photometric calibration we scale the outer part of our profile to the  profile from \cite{Trager}, using their Chebychev fit, and extend it with the outer points of this profile.
\begin{figure}
\resizebox{\hsize}{!}{\includegraphics{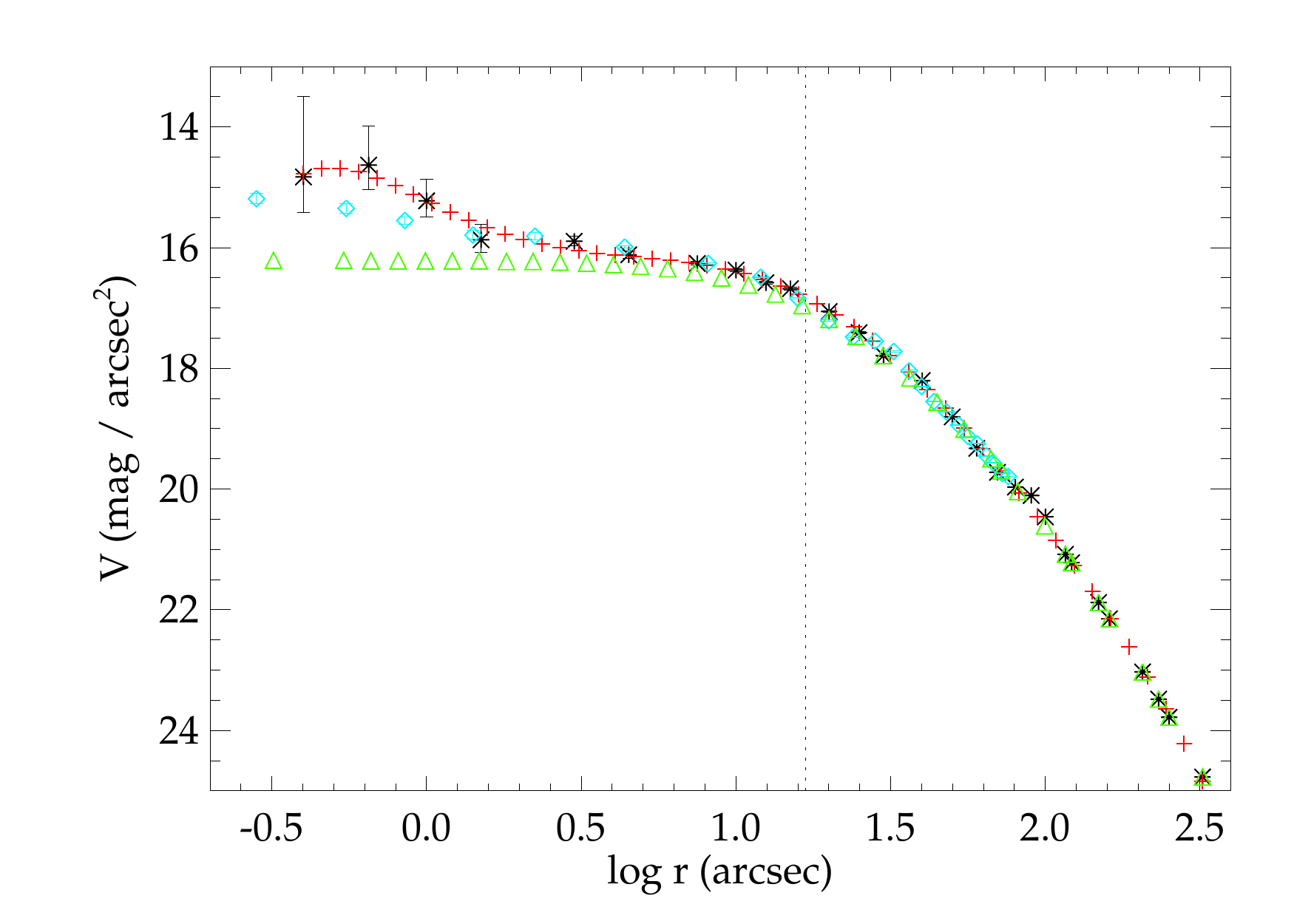}}
\caption{Comparison of the four surface brightness profiles used for later analysis: Our combined profile (black asterisks), the smoothed  profile obtained from a fit with Chebychev polynomials (red plus signs), the profile from \cite{NG06} (cyan diamonds), and the \cite{Trager} profile (green triangles,  Chebychev fit on photometric points). The dotted vertical line denotes the core radius of 16.8\arcsec\, \citep{harris}.}
\label{fig:sbcall}
\end{figure}
\begin{table}
\caption{The combined surface brightness profile in the V-band. $\Delta$V$_{l}$ and $\Delta$V$_{h}$ are the low and high values of the errors.}
\label{tab:sbprof}
\centering
\begin{tabular}{c c c c}
\noalign{\smallskip}
\hline\hline
\noalign{\smallskip}
log r &V &$\Delta$V$_{l}$ &$\Delta$V$_{h}$ \\
$ \left[ \mbox{arcsec} \right]$  & $\left[ \mbox{mag/arcsec}^{2} \right]$  &$ \left[ \mbox{mag/arcsec}^{2} \right]$  &$ \left[ \mbox{mag/arcsec}^{2} \right]$  \\
\noalign{\smallskip}
\hline
\noalign{\smallskip}
$-0.40$ & $14.83$ & $1.34$ & $0.59$ \\
$-0.19$ & $14.63$ & $0.65$ & $0.41$ \\
$0.00$ & $15.22$ & $0.36$ & $0.27$ \\
$0.18$ & $15.87$ & $0.26$ & $0.21$ \\
$0.48$ & $15.90$ & $0.08$ & $0.08$ \\
$0.65$ & $16.12$ & $0.06$ & $0.06$ \\
$0.88$ & $16.26$ & $0.03$ & $0.03$ \\
$1.00$ & $16.37$ & $0.03$ & $0.03$ \\
$1.10$ & $16.58$ & $0.03$ & $0.03$ \\
$1.18$ & $16.68$ & $0.02$ & $0.02$ \\
$1.30$ & $17.06$ & $0.01$ & $0.01$ \\
$1.40$ & $17.41$ & $0.01$ & $0.01$ \\
$1.48$ & $17.79$ & $0.01$ & $0.01$ \\
$1.60$ & $18.20$ & $0.01$ & $0.01$ \\
$1.70$ & $18.80$ & $0.01$ & $0.01$ \\
$1.78$ & $19.32$ & $0.01$ & $0.01$ \\
$1.85$ & $19.72$ & $0.01$ & $0.01$ \\
$1.90$ & $19.97$ & $0.01$ & $0.01$ \\
$1.95$ & $20.11$ & $0.01$ & $0.01$ \\
$2.00$ & $20.46$ & $0.01$ & $0.01$ \\
$2.07$ & $21.08$ & $0.01$ & $0.01$ \\
$2.09$ & $21.22$ & $0.01$ & $0.01$ \\
$2.17$ & $21.88$ & $0.01$ & $0.01$ \\
$2.21$ & $22.15$ & $0.01$ & $0.01$ \\
$2.31$ & $23.03$ & $0.01$ & $0.01$ \\
$2.37$ & $23.48$ & $0.01$ & $0.01$ \\
$2.40$ & $23.78$ & $0.01$ & $0.01$ \\
$2.51$ & $24.77$ & $0.01$ & $0.01$ \\
\noalign{\smallskip}
\hline
\noalign{\smallskip}
\end{tabular}
\end{table}
In Figure \ref{fig:sbcall} our adopted surface brightness profile is indicated by  asterisks and compared to the profiles from \citealt{NG06} (diamonds), and from  \citealt{Trager} (triangles). The \cite{Trager} profile is fainter and flat towards the center. This profile was derived from from ground based images, and  the inner region of the cluster is not as well spatially resolved as with data from space.  Our profile is the brightest, but due to the method and the data we use, the inner points of the profile have a high uncertainty, which comes from the Poisson statistics of the number of stars in each bin.  The exact values of our profile are listed in Table \ref{tab:sbprof}.
However, for later analysis we  also use a smoothed profile. This is obtained by fitting Chebychev polynomials to our derived profile, as indicated by  plus signs in Figure \ref{fig:sbcall}.
We further check whether the difference to the surface brightness profile from \cite{NG06} comes from the different center position or the different methods. So we compute a profile with our method, but using the \cite{NG06} center. At large radii the results for the different centers are very similar, but in the central part the difference becomes larger, but not more than 0.35 mag/arcsec${^2}$. The error bars are so large that it is difficult to  say whether the choice of center influences the shape of the profile. But  the two surface brightness profiles with the distinct centers  both tend to be higher than the \cite{NG06} profile. So the  difference with their profile seems to come mostly from the different methods used to compute the surface brightness, and not from the different centers.

\section{Spectroscopy}
\label{sec:spec}
\subsection{Observations}

\begin{figure}
\resizebox{\hsize}{!}{\includegraphics{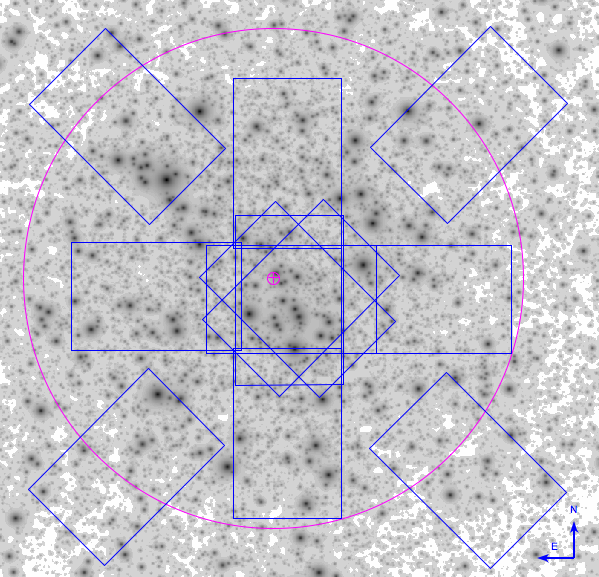} }
\caption{The positions of the ARGUS pointings are indicated  as rectangles. The  circle indicates the core radius of \cite{harris}, which is 16.8\arcsec. The underlying image is the synthetic I-band image from the HST star catalog (see Section \ref{sec:velmap}).}
\label{fig:vltexp}
\end{figure}

The spectroscopic data were obtained during two nights with the GIRAFFE spectrograph of the FLAMES instrument  \citep{flames} on 2010-05-05 and 2010-05-06 at the Very Large Telescope (VLT). Observations were made in  ARGUS mode with 1:1 magnification scale, providing a total coverage of 11.5\arcsec\,x 7.3\arcsec, and a sampling of 0.52\arcsec\,per microlens. We used the low resolution grating LR8, which covers the wavelength range from 8206 to 9400\,\AA\, and has a resolving power R $\approx$ 10,400. At  the wavelengths 8498\,\AA, 8542\,\AA, and 8662\,\AA\,are the Calcium triplet lines,  which are known to be well suited for kinematic analysis. 
Old stars, like those present in globular clusters, show strong features from these lines.
We obtained pointings of the cluster's center and of adjacent  regions around the core radius, every pointing has three exposures of 600\,s each. Figure \ref{fig:vltexp} illustrates the arrangement of the pointings.

\begin{figure*}
\centering
\includegraphics[width=17cm]{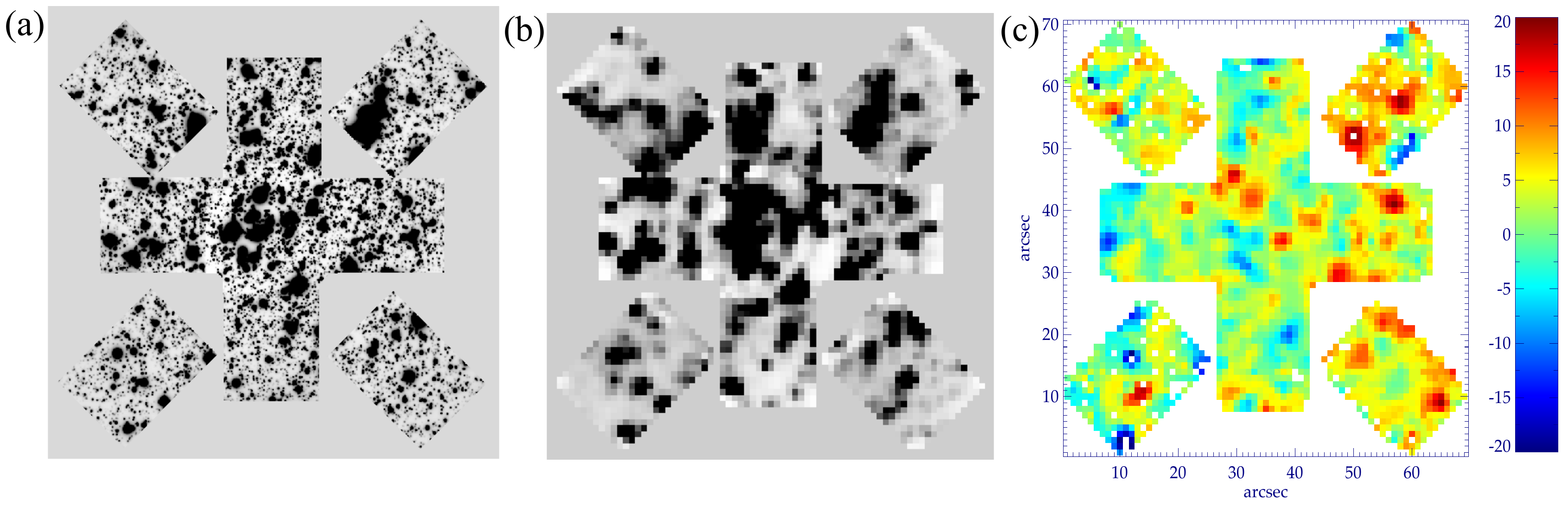} 
\caption{(a) The  synthetic image cut to the ARGUS pointings, (b) the reconstructed ARGUS image, and (c) the velocity map. Colors are as indicated by the color bar in units of km/s. The blue spaxels of the velocity map indicate approaching stars, the red spaxels are receding stars.  White spaxels mean that within this spaxel  no spectrum was found that could be used to determine the fit parameters. North is  right, East is up.}
\label{fig:combined}
\end{figure*}  

\subsection{Data reduction}
Data reduction includes removing cosmic rays, pipeline reductions, and sky subtraction. We follow the procedure  largely developed in \cite{Nora}, except for a change in the order of the steps. At the beginning, we use the \emph{Laplacian Cosmic Ray Identification L.A.Cosmic} written by \cite{lacosmic} to remove cosmic rays from the raw images.
The GIRAFFE pipeline, programmed by ESO, consists of five recipes: In a first step, \emph{gimasterbias} creates a masterbias and a bad pixel map from a set of raw bias frames. The recipe \emph{gimasterdark}   corrects every raw dark frame for the bias, and creates a masterdark using the median. \emph{Gimasterflat} creates a masterflat, which contains information on pixel-to-pixel variations and fiber-to-fiber transmissions. This recipe also locates the position and width of the spectra on the CCD from the flat-field lamp images. It detects and traces the fibers on the masterflat and records the location and width of the detected spectra.
The recipe \emph{giwavecalibration} uses a ThAr arc lamp calibration frame to compute the dispersion solution and the setup specific slit geometry table. The spectrum from  the ThAr lamp frame is extracted using the fiber localization (provided by \emph{gimasterflat}). The final pipeline recipe is \emph{giscience}, which applies the calibrations produced by the other recipes to the cosmic ray removed object frames. As a result we obtain bias corrected, dark subtracted, flat fielded, extracted and wavelength calibrated spectra, which are also corrected for fiber-to-fiber transmission and pixel-to-pixel variations.
 The next step of data reduction is sky subtraction.  Mike Irwin developed a program, which is described in  \cite{skysub}, and our program is based on this approach. It splits the sky spectra in continuum and sky lines, and subtracts both from the object spectra.    
We also normalize our spectra. As we are solely interested in the kinematic extraction from the spectra, we do not perform flux calibration.

 \section{Kinematics}
\label{sec:kin}
The spectroscopic data contain information on the central kinematics of NGC 5286, and a velocity-dispersion profile is essential for our models. This section describes the construction of a velocity map, and of a velocity-dispersion profile. For the inner part of NGC 5286 we use the FLAMES data, for the outer part we use  data from the Rutgers Fabry Perot (RFP).

\subsection{Velocity map}
\label{sec:velmap}
\begin{figure}
\resizebox{\hsize}{!}{\includegraphics{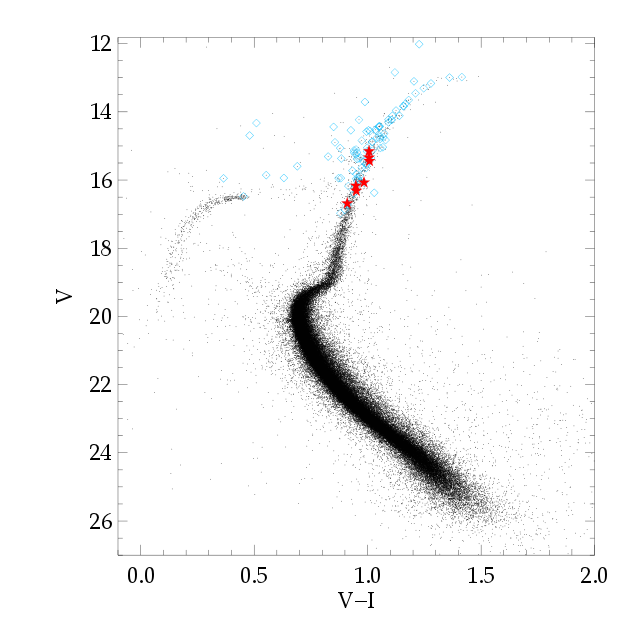} }
\caption{Color-magnitude diagram from the photometric HST star catalog of stars with low uncertainty in photometry. The star symbols denote the stars used for the  template spectrum,  diamonds denote stars that contribute more than 70\% of the light to a spaxel.}
\label{fig:cmd}
\end{figure}
We match  the reconstructed images of ARGUS produced by the \emph{giscience} recipe  to a synthetic  image in the I-band, constructed from the photometric star catalog. Thus we identify which region the FLAMES data cover exactly with respect to the photometric center. We  use the synthetic image since  there is an overexposed region in the HST V-band image, so that the program does not adjust the north-eastern ARGUS images properly. At first, the angle between the synthetic image and each ARGUS pointing is evaluated,  and the synthetic image is rotated by this angle. Then the synthetic image is  blurred by convolving the image with  the point spread function  that is obtained from a Gaussian function with  $\sigma$ = 0.6\arcsec, roughly corresponding to the seeing during the spectroscopic observations. We then  cross correlate the ARGUS pointings with the respective blurred synthetic image. This procedure is done iteratively, until the offset between the ARGUS and synthetic images is smaller than 0.1 ARGUS pixel, which is achieved after four iterations at most. So, every spectrum is correlated to a  certain coordinate, and we obtain  an area that covers 69 x 70 spaxels. Figure \ref{fig:combined} is an illustration of  (a) the synthetic image, cut to match  the combination of ARGUS pointings, (b) the  reconstructed ARGUS pointings, and (c) the velocity map, the colors denote different values as indicated by the color bars. 

Every spectrum  is a sum of individual stellar spectra, weighted by the luminosities of the stars. To extract the velocity and velocity dispersion from the integrated spectrum one needs a spectral template. To find a good template we use the photometric star catalog with information on the positions of the stars. The number of stars that contribute to one spaxel,  
and the fraction of light coming from a certain star in the spaxel, are calculated. 
So we can identify  stars that dominate a single spaxel. Figure \ref{fig:cmd}  shows the color-magnitude diagram from the HST catalog, where stars that dominate a spaxel by at least 70\% are marked as blue diamonds. The star symbols denote seven  stars, which contribute at least 84\% of the light in one spaxel and are fainter than m$_V$=15.  According to their position on the CMD, these stars are most probably cluster members. Those stars' spectra were combined to a master template spectrum. We found the spectral lines of brighter stars are less Gaussian, possibly due to saturation, and therefore not suitable as template spectra. 

We use the penalized pixel-fitting (\emph{pPXF}) program developed by \cite{ppxf} to derive a velocity and velocity dispersion for each spaxel. 
As we have three exposures for every pointing and some pointings overlap, there can be several spectra for one spaxel. Therefore these spectra are identified, and sigma clipping is applied about the median of the data set for every spaxel, and we use the mean of the spectra for further analysis.  It is possible that only one or a few bright stars dominate the light in their environment. 
Therefore we check in which  spaxel either  a single star contributes more than 70\% of the light, or  less than five stars contribute to one spaxel.  578 spaxels of 3237 are affected by  shot noise.  Those spaxels are excluded from  further analysis.

\subsection{Inner velocity-dispersion profile}
\label{sec:sigmain}
To determine the velocity-dispersion profile, we follow the procedure  described in \cite{Nora2808} and construct circular bins around the center of the cluster. We choose seven independent, non-overlapping circular bins. A higher number of bins results in low signal-to-noise. The  spaxels that are affected by shot noise are excluded. We  check different combinations of minimum star number and light contribution, and different angular bins, but the overall shape of the kinematic profile does  not change much. 
In each circular bin the spectra are combined and we    apply the \emph{pPXF} method to compute the  moments of the line-of-sight velocity distribution (LOSVD). As we combine all the spectra around the center in circular bins, we obtain the mean velocity of the circular bin $\langle V \rangle_{bin}$, and $V_{rms}=\sqrt{V_{rot}^{2}+\sigma^{2}}$, with the rotational velocity $ V_{rot} $ and the velocity dispersion  $ \sigma $.  $V_{rms}$ is  required for the  models, and we will refer to the $V_{rms}$ profile as velocity-dispersion profile. 
However, the Gauss-Hermite moments, which are the output of the \emph{pPXF} program, need to be corrected as described in \cite{linecorrect}. Depending on the number of fitted Gauss-Hermite moments, we obtain different results. We fit  the first two  $(V, \sigma)$, four $(V, \sigma, h_3, h_4)$, or six $(V, \sigma, h_3, h_4, h_5, h_6)$ moments. 

As alternative approach we calculate the velocity-dispersion profile with a non-parametric fit (see \citealt{nonpar, Pinkney}). The observed cluster spectrum is deconvolved using the template spectrum. Deconvolution is done using a maximum penalized likelihood (MPL) method, and one obtains a non-parametric LOSVD. For the fitting procedure, the first step is to choose an initial velocity profile in bins. This profile is convolved with the template to obtain a cluster spectrum, which then  is used to calculate the residuals to the observed spectrum. The parameters for the LOSVD are varied to obtain the best fit. 

The error for $V_{rms}$  comes from Monte Carlo simulations, and  is an estimate for the amount of shot noise. 
For every bin the kinematics are calculated in 1,000 realizations. The shot noise error is the sigma-clipped standard deviation of these 1,000 velocity dispersions. 
The uncertainties for the other moments of the LOSVD and for $V_{rms}$ of the \emph{pPXF} method are obtained from Monte Carlo simulations on the spectra. As recommended in \cite{ppxf}, we repeat the measurement process for a large number of different realizations of the data. This is done by adding noise to the original binned cluster spectra. After 100 realizations we use the standard deviation of the kinematic parameters as the corresponding uncertainties.

\begin{figure}
\resizebox{\hsize}{!}{\includegraphics{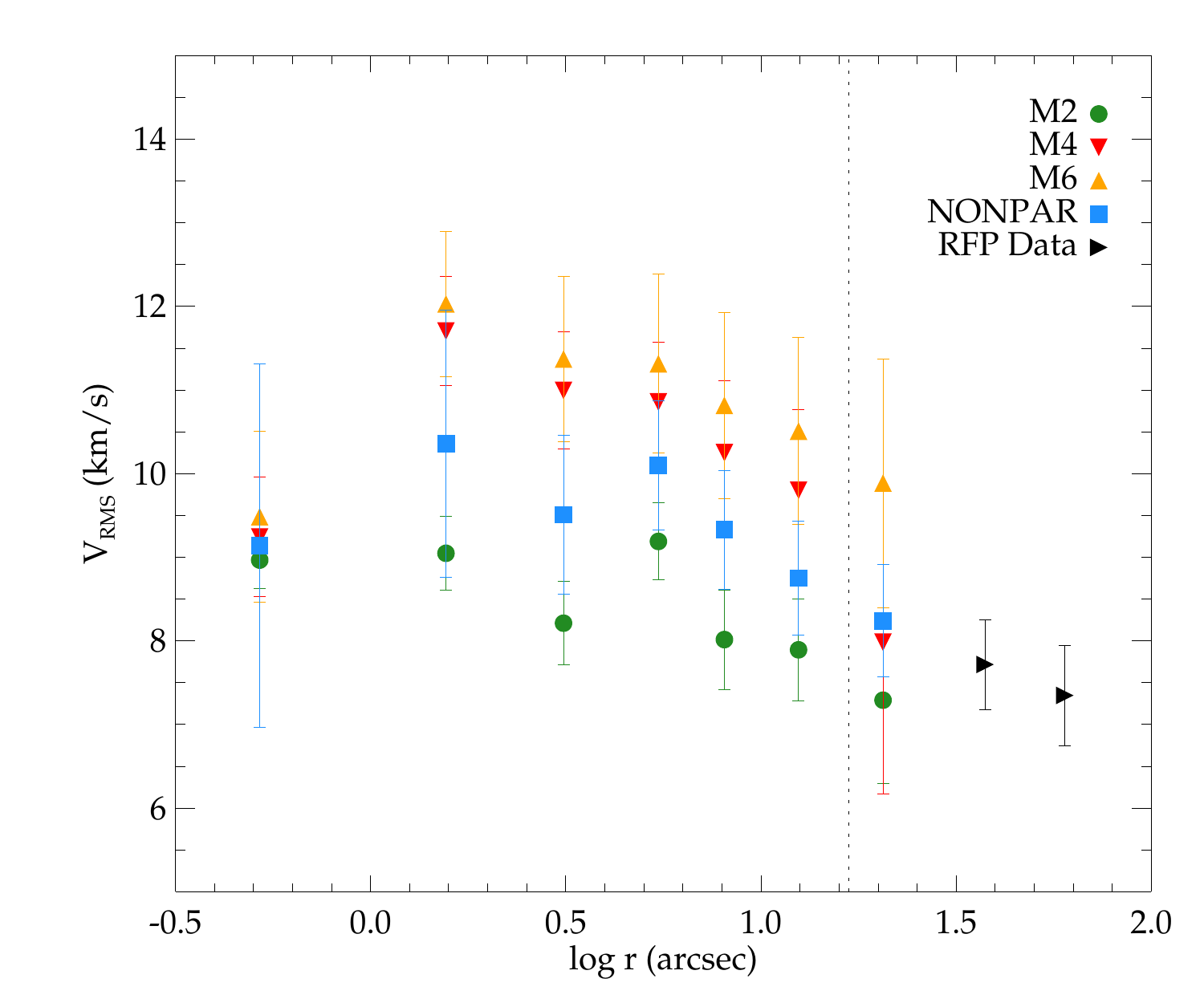} }
\caption{Velocity-dispersion profiles  with error bars. Different symbols indicate different methods of calculation. M2, M4, and M6 are the profiles computed with \emph{pPXF}, fitting only the first  2, 4, or 6 moments, respectively. The  squares denote the profile calculated with the nonparametric method. The triangles are from outer kinematic data, and the dotted vertical line denotes the core radius at 16.8\arcsec\,\citep{harris}.}
\label{fig:sigprofvgl}
\end{figure}
Figure \ref{fig:sigprofvgl} illustrates the outcome of the different methods. M2 is the \emph{pPXF} velocity-dispersion profile when only the first two parameters ($V, \sigma$) are fitted, M4 when four parameters  ($V, \sigma, h_3, h_4$) are fitted, and M6 when six parameters ($V, \sigma, h_3, h_4, h_5, h_6$) are fitted. The velocity dispersion values calculated with the non-parametric method lie in between M2 and M4, and the shape of the velocity-dispersion profiles are all quite similar, but the absolute values computed with  \emph{pPXF} are very sensitive to the number of calculated moments. Therefore we use the non-parametric profile for further analysis. It is remarkable that the difference between M2, M4, and M6 is less than 1 km/s for the innermost bin, but up to 3 km/s for the other bins. We tried to shift the template spectrum by the template velocity, or use another template spectrum, but the discrepancy in the velocity dispersion remains.
\begin{table}
\caption{The  kinematics of NGC 5286  obtained from  FLAMES and RFP data. For the FLAMES data, $\langle V \rangle_{bin}$ is the velocity relative to the template velocity.}
 \label{tab:sigprof}
 \centering
\begin{tabular}{cccccc}
\noalign{\smallskip}
\hline\hline
\noalign{\smallskip}
\multicolumn{6}{c}{FLAMES DATA}\\
\noalign{\smallskip}
\hline
\noalign{\smallskip}
$ r $&$\langle V \rangle_{bin}$&$\Delta\langle V \rangle_{bin}$&$V_{rms}$&$\Delta V_{rms}$&$S/N$\\
$ \left[  arcsec \right]  $&$ \left[  km/s \right]  $& $ \left[  km/s \right]  $ &$ \left[  km/s \right] $& $ \left[  km/s \right]  $ &\\
\noalign{\smallskip}
\hline
\noalign{\smallskip}
$0.52$ & $2.4$&  $0.4$ & $9.1$&  $2.2$ & $  112$ \\
$1.56$ & $4.6$&  $0.4$ & $10.4$&  $1.6$ & $  133$ \\
$3.12$ & $2.7$& $0.3$ & $9.5$&  $1.0$ & $  140$ \\
$5.46$ & $2.1$&  $0.4$ & $10.1$&  $0.8$ & $  142$ \\
$8.06$ & $2.7$&  $0.3$ & $9.3$&  $0.7$ & $  138$ \\
$12.48$ & $2.4$&  $0.4$ & $8.8$&  $0.7$ & $  134$ \\
$20.54$ & $2.1$&  $0.4$ & $8.2$&  $0.7$ & $   97$ \\
\noalign{\smallskip}
\hline\hline
\noalign{\smallskip}
\multicolumn{6}{c}{RFP DATA}\\
\noalign{\smallskip}
\hline
\noalign{\smallskip}
$ r $&$\langle V \rangle_{bin}$&$\Delta\langle V \rangle_{bin}$&$V_{rms}$&$\Delta V_{rms}$&$Number$\\
$ \left[  arcsec \right]  $&$ \left[  km/s \right]  $& $ \left[  km/s \right]  $ &$ \left[  km/s \right] $& $ \left[  km/s \right]  $ &$of\,stars$\\
\noalign{\smallskip}
\hline
\noalign{\smallskip}
$37.50$ &$59.0$& $0.8$& $7.7$&$0.5$ & $109$ \\ 
$60.00$ &$57.9$& $0.8$& $7.3$&$0.6$ & $80$ \\
\noalign{\smallskip}
  \hline
\noalign{\smallskip}
\end{tabular}
\end{table}
Table \ref{tab:sigprof} records the results of the  inner kinematic measurements with the non-parametric method. The first column denotes the  radii of the bins. The following columns list the kinematic parameters, which are the velocities of each bin $\langle V \rangle_{bin}$ in the cluster reference frame relative to the template spectrum, and the second moment $V_{rms}$ with respective errors in units of km/s.  The signal-to-noise ratio S/N is higher than 112 except for the outermost bin, where S/N=97.

In order to  estimate the radial velocity of the cluster in the heliocentric reference frame, we combine all spectra in the pointings and measure the velocity relative to the velocity of the template. The velocity of the template $V_{r, temp}$ results in (56.9 $\pm$  0.6)\,km/s, 
and the velocity of all combined spectra relative to it is  $V_{r, cluster}$~=~(2.4 $\pm$ 0.6)\,km/s. 
 So the heliocentric velocity of the cluster $V_{r}$ =  $V_{r, temp}$ + $V_{r, cluster}$ =~(59.3 $\pm$ 1.2)\,km/s. This is in agreement with the value from \cite{harris},  $V_{r}$ =~(57.4 $\pm$ 1.5)\,km/s.

\subsection{Outer velocity-dispersion profile}
\label{sec:sigmaout}
In addition to spectroscopic data in the center we  use a second data set for larger radii, i.e. up to r\,$\sim$\,75\arcsec. The data were obtained with the 4-m Victor M. Blanco Telescope at the Cerro Tololo Inter-American Observatory (CTIO). Observations were made from 1994-05-31 to 1994-06-03, 
 with the Rutgers Fabry Perot (RFP).  The used methods of observation and data reduction are similar to those described in \cite{fpdata} for M15.  The data set for NGC 5286 contains 1,165 velocities of individual stars with errors, flux, and information about their position relative to a center in arcsec. With an RFP image of the pointing, the HST image, and the \emph{iraf} recipes \emph{daofind}, \emph{ccmap}, and \emph{xyxymatch}, we transform the coordinates to the world coordinate system (WCS). 
 
 To determine the velocity dispersion we draw circular bins around the center and use only stars with low error in velocity and with a high flux. This is done to make sure that only stars with a well-determined velocity are used for further analysis. Crowding in the cluster's center is known to cause problems in the determination of velocity. Background light contaminates the measurements for fainter stars, and this biases the velocities of the stars to the mean velocity of the cluster. Therefore, the velocities obtained from Fabry-Perot measurements are often too low. We use only stars in the outer part of the cluster, where the crowding effect is less relevant. With the maximum likelihood method introduced by \cite{maxlik} we calculate the velocity dispersion $V_{rms}$ and mean velocity $\langle V \rangle_{bin}$. 
 The obtained values are listed in Table \ref{tab:sigprof}. The last column displays the number of stars that were used to determine $V_{rms}$ for each bin. Both values of $\langle V \rangle_{bin}$ are in agreement with the heliocentric velocity of the cluster from \cite{harris},  $V_{r}$ = (57.4 $\pm$ 1.5) km/s.
 
 We also compute the effective velocity dispersion $\sigma_{e}$, as defined in  \cite{gueltekin}
\begin{equation}
\sigma_e^2=\frac{\int^{R_e}_0 V_{rms}^2 I(r) dr}{\int^{R_e}_0  I(r) dr},
\end{equation}
where R$_e$ is the effective radius or half-light radius (43.8\arcsec, \citealt{harris}), I(r) is the surface brightness profile, and V$_{rms}$ is the velocity-dispersion profile. We obtain $\sigma_{e}$\,=\,(9.3 $\pm$ 0.4)\,km/s, this is higher than the central velocity dispersion of (8.1 $\pm$ 1.0) km/s from \cite{harris}, but the values agree within their uncertainty limits.

 \subsection{Rotation}
 
\begin{figure}
	\resizebox{\hsize}{!}{\includegraphics{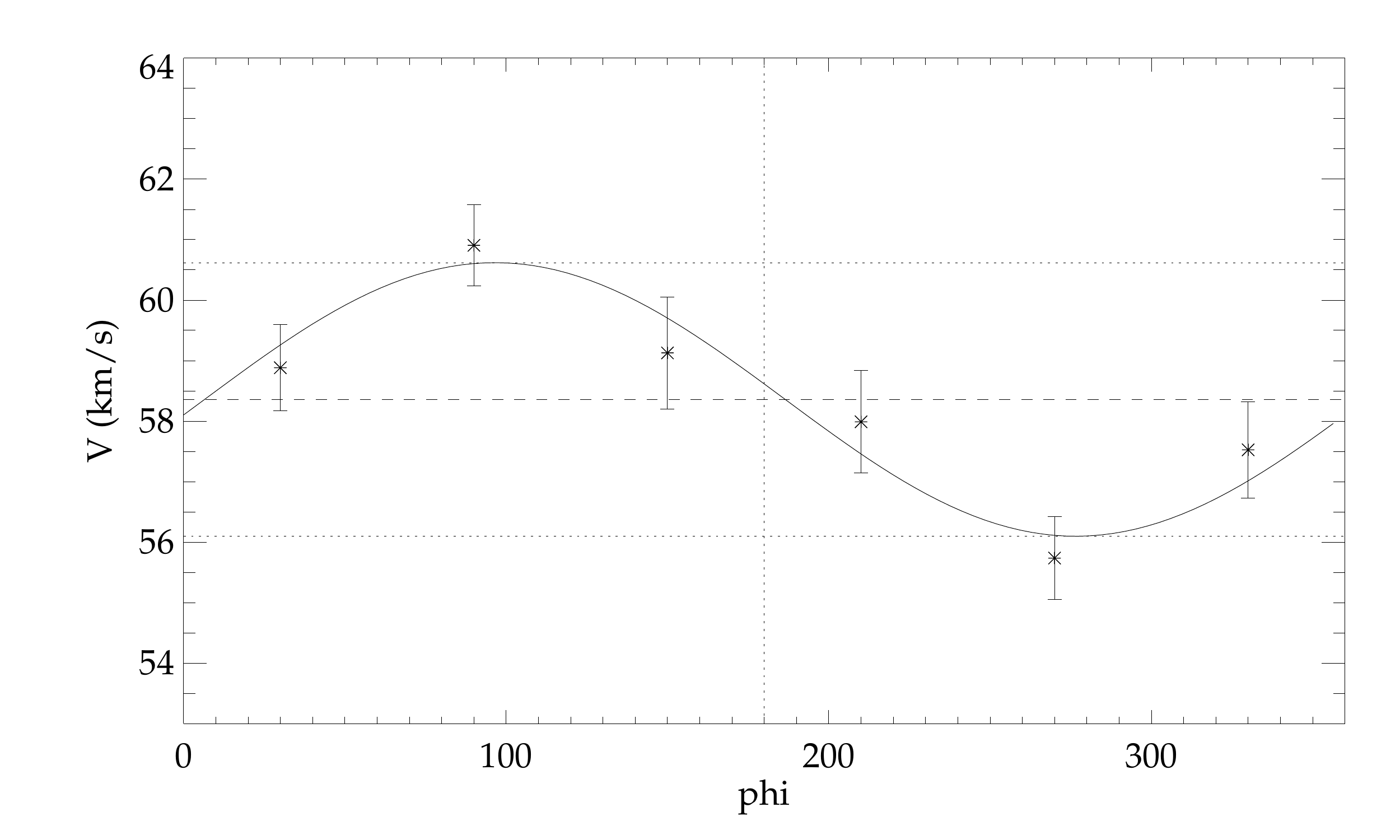} }
	\caption{The mean radial velocities with error bars of stars at a distance r $\geq$ 25\arcsec\, plotted versus position angle $\phi$ for 6 segments from RFP data. The solid line is the fit to the data.}
	\label{fig:phaseplot}
	\end{figure}

 Some globular clusters are known to rotate, for example $\omega$\,Centauri and 47 Tucanae (\citealt{omcenrot}).  
As NGC 5286  is not totally spherical and has an ellipticity $\epsilon$ = 0.12, comparable to the ellipticity of $\omega$\,Centauri ($\epsilon$ = 0.17) and 47 Tucanae ($\epsilon$ = 0.09,  all ellipticity values are from \citealt{white}), we investigate a possible rotation of NGC 5286 with our two data sets.  
For the outer part of the cluster we use the RFP data. We separate the cluster in 6 to 18 segments and find all stars with a distance from the cluster center of at least 25\arcsec. The mean velocity of the stars in these areas is determined with the  maximum likelihood method. In Figure \ref{fig:phaseplot} the mean radial velocity V of the stars in the segments is plotted against the position angle $\phi$. For six segments there are more than 85 stars in each bin, for 18 segments at least 20 stars. We fit the function $f$
	\begin{equation}
	f=A\sin(\phi + \psi) + V_r,
	\end{equation}	
with the parameters amplitude $A$, phase angle $\psi$, and heliocentric cluster velocity $V_r$ to the data points for all segment numbers from 6 to 18. The best-fit parameters  are averaged, and we obtain for the mean heliocentric velocity  $V_r$ = (58.3 $\pm$ 1.1) km/s, and for the amplitude A = (2.3 $\pm$ 1.5) km/s. The phase $\psi$ is  (\mbox{-0.01}\,$\pm$\,0.7)\,rad,
and the position angle of the rotation axis is (89 $\pm$ 40)\degr, measured positive from north to east. \cite{white} found the orientation of the major axis at $\theta$=21\degr.

The spectroscopic FLAMES data is used to investigate  indications of rotation in the inner part of the cluster. In some  segments the spectra have low signal-to-noise and therefore the mean radial velocities have large error bars of up to more than 3\,km/s.  For the sine fit we use only segments with a signal-to-noise higher than 40. However, the result for the phase angle $\psi$ depends highly on the number of fitted segments, and therefore we  find no significant indication for rotation in the inner 25.5\arcsec of NGC 5286.

 \section{Jeans models}
 \label{sec:jeans}
\begin{figure}
\resizebox{\hsize}{!}{\includegraphics{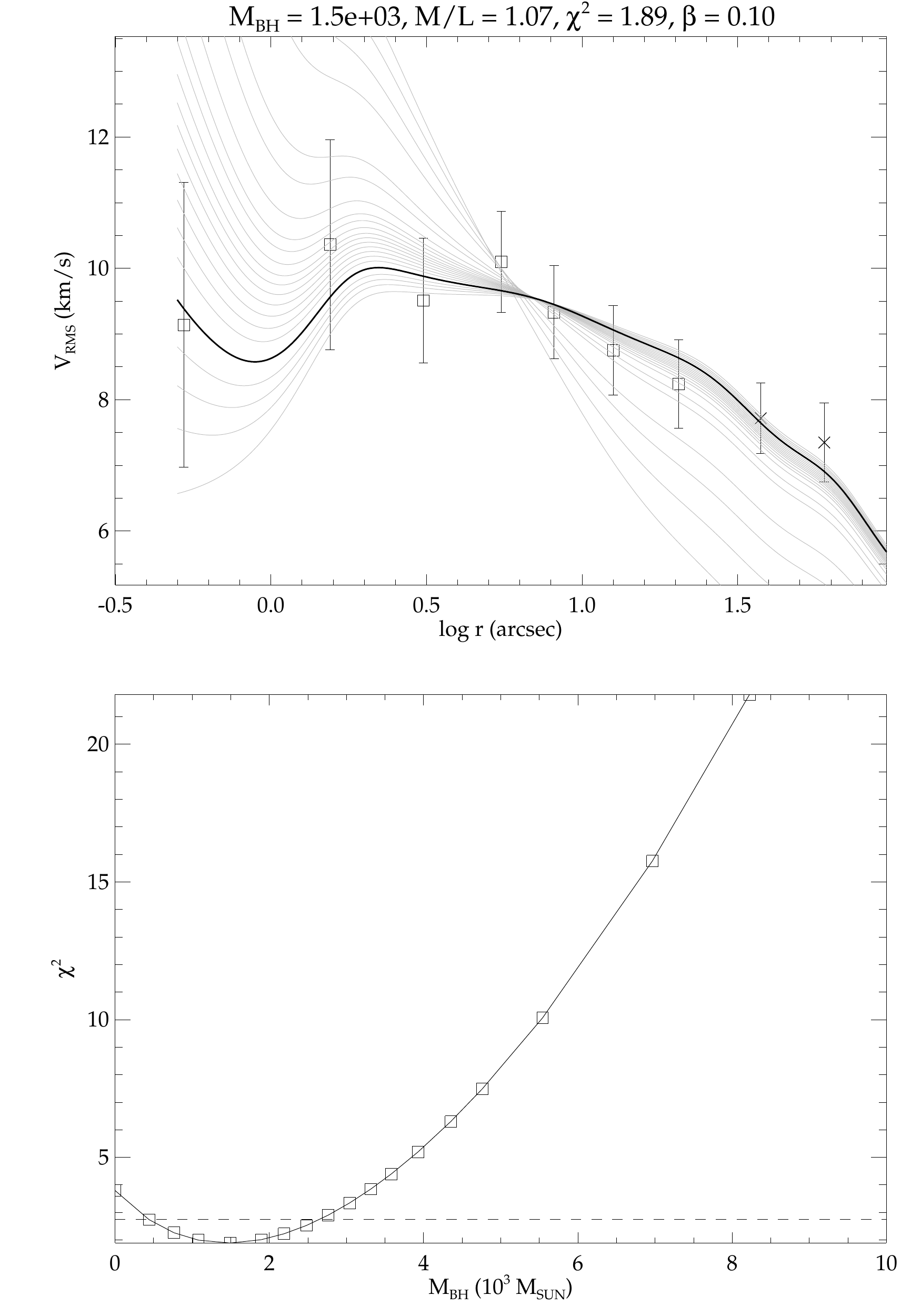} }
\caption{Different spherical Jeans models fitted to our combined surface brightness profile  for fixed anisotropy $\beta$=0.1. The upper panel shows the models together with the data for different black-hole masses, the thick black line indicates the best fit.  The lower panel shows the $\chi^2$ values as a function of black-hole mass, and the dashed line denotes $\Delta \chi^2$ = 1. The best fit with $\chi^2$ = 1.89 requires   a black-hole mass of M$_\bullet$\,=\,($1.5^{+1.0}_{-1.1}$)\,$\cdot$\,10$^3$\,M$_\odot$. }
\label{fig:sphcomb}
\end{figure}
Jeans models provide an analytical approach to fit kinematics of the globular cluster NGC~5286, and we use an \emph{IDL} routine developed by 
\cite{JAM}. In order to solve the Jeans equations,  the mass density $\varrho$ is required as an input. We use the surface brightness profile for NGC 5286 derived in Section \ref{sec:SB}, and deproject it to obtain the luminosity density $j$ of the cluster, which is related to the mass density $\varrho$ by the mass-to-light ratio (M/L).   The surface brightness profile is deprojected using the Multi-Gaussian expansion (MGE) method, developed by \cite{ericmge}, using the \emph{IDL} routine  written by \cite{MGE}. With the MGE formalism there are Jeans equations for every single Gaussian component, and after solving the equations, we obtain the second velocity moment $\overline{v_{los}^{2}}$(R) along the line-of-sight.  The results are  fitted to the observed velocity-dispersion data points, as computed in Sections \ref{sec:sigmain} and  \ref{sec:sigmaout}. 
It is  possible to vary the values of anisotropy $\beta$\,=\,1\,$-$\,$\sigma_{\theta}^{2} / \sigma_{R}^{2}$ and M/L for every individual MGE Gaussian.
 The velocity-dispersion profile can be scaled to the observations by adjusting M/L. Therefore all models meet in one point at which the model is scaled.

\subsection{Spherical Jeans models with constant M/L}

\begin{table*}
\caption{The spherical Jeans model best-fit values with constant M/L.}
 \label{tab:sphe}
 \centering
\begin{tabular}{lcccccc}
\noalign{\smallskip}
\hline\hline
\noalign{\smallskip}
 surface brightness profile&	$\beta$&	$M/L$&	$M_\bullet$&	$M_{tot}$&	$L_{tot}$&	$\chi^2$\\
& constant	& $ \left[  M_\odot/L_{V,\odot}  \right]  $ &	$ \left[  10^3\,M_\odot \right] $& 	$ \left[  10^5\,M_\odot \right]  $ &$\left[  10^5\,L_\odot \right]  $&\\

\noalign{\smallskip}
\hline
\noalign{\smallskip}
combined profile& $0.10$&  $1.07\pm0.03$ & $1.5^{+1.0}_{-1.1}$&  $2.89\pm0.10$ & $ 2.71 ^{+0.05}_{-0.04}$&$1.89$ \\
&&&&&&\\
smooth profile& $0.05$&  $1.07 ^{+0.04}_{-0.03}$ & $1.9 ^{+0.9}_{-1.2}$&  $2.85 \pm0.12$ & $2.66 ^{+0.05}_{-0.04}$&$2.28$ \\
&&&&&&\\
\cite{NG06} & $0.10$& $1.11 ^{+0.04}_{-0.05}$ & $1.1  ^{+1.4}_{-1.1}$&  $2.55 \pm0.12$ & $ 2.30 ^{+0.03}_{-0.02}$&$2.34$ \\
&&&&&&\\
\cite{Trager} & $0.15$&  $1.22 ^{+0.00}_{-0.06}$ & $0.0 ^{+1.6}_{-0.0}$&  $3.08 ^{+0.00}_{-0.14}$ & $2.51$&$2.22$ \\

\noalign{\smallskip}
  \hline
\noalign{\smallskip}
\end{tabular}
\end{table*}
The simplest model is a spherical Jeans model. We consider  different values of anisotropy $\beta$, and different surface brightness profiles. The profiles used are the Chebychev fit from \citealt{Trager},  \citealt{NG06},  our derived profile from a combination of star counts and integrated light measurements (see Table \ref{tab:sbprof},  combined profile), and the profile from a fit of Chebychev polynomials to the combined profile to make it smooth (smooth profile). The best fits are obtained with our combined  profile. For each profile we test different values of constant anisotropy $\beta$ in the interval of [-0.2, 0.3] over the entire cluster radius. We fit a global M/L, and the best fit is obtained with radial anisotropy $\beta$=0.1, M$_\bullet$=1.5$\cdot$10$^3$\,M$_\odot$, and constant M/L=1.07 (all values of M/L are in solar units of M$_\odot$/L$_{V,\odot}$). The outcome  is shown in Figure \ref{fig:sphcomb}. For the other profiles the best fit values are listed in Table \ref{tab:sphe} with errors from the $\Delta \chi^2$=1 limit. \cite{Trager} do not provide uncertainty limits to their surface brightness profile, therefore the uncertainty in M$_{tot}$ comes only from the uncertainty in M/L. The value from \cite{harris} for the integrated V-band magnitude is 7.34, this corresponds to a total luminosity L$_{tot}$ of 2.40\,$\cdot$\,10$^5$\,L$_\odot$. The \cite{Trager} profile has too low resolution at the center and is therefore not as suitable for our analysis as the other profiles. 

\begin{figure*}
	\centering
	\includegraphics[width=17cm]{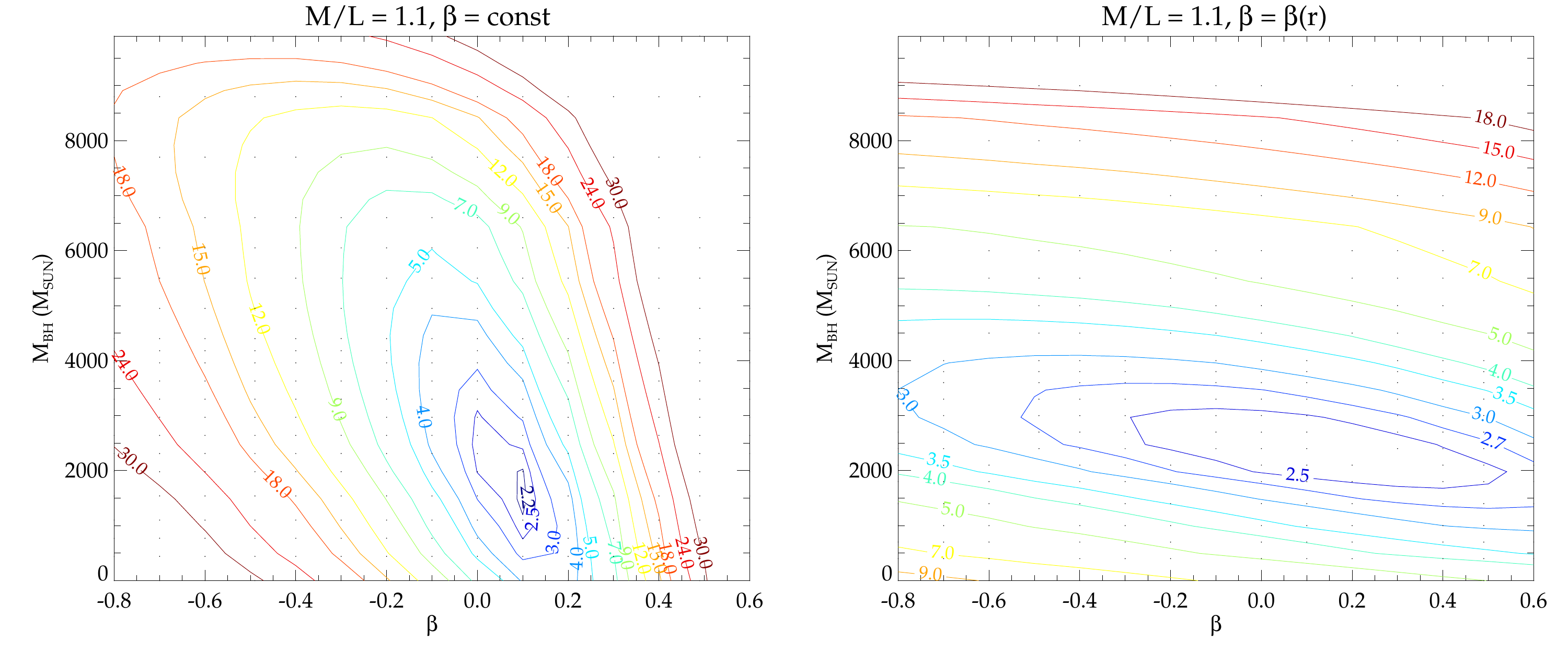}
	\caption{$\chi^2$ contours describing the agreement between the data and spherical Jeans models of the combined profile with  constant mass-to-light ratio M/L=1.1, the numbers denote the respective value of $\chi^2$.  Left: $\beta$ = const., right: $\beta \propto$ r. In the case of radially varying anisotropy the abscissa denotes the value of $\beta_{out}$, the maximum anisotropy in the outermost part of the cluster.}
	\label{fig:sphcont}
	\end{figure*} 
We also assume a radial shape of anisotropy $\beta \propto$ r, with isotropy in the very center of the cluster, and rising anisotropy towards the outskirts. At the outermost Gaussian the anisotropy $\beta_{out}$ is highest. This approach is justified as the relaxation time is short in the cluster's center (t$_{relax}\approx$ 2.5\,$\cdot$10$^8$\,yr, \citealt{harris}), and higher at larger radii (at the half-mass radius $t_{relax}$ $\approx\,$ 1.3\,$\,\cdot\,10^{9}$\,yr, \citealt{harris}). \cite{Nora} ran an $N$-body simulation of a globular cluster with initial anisotropy and  showed that  it becomes more isotropic after few relaxation times. Assuming an age of 11.7\,$\cdot$10$^9$\,yr \citep{absage},  NGC\,5286 is already 46 relaxation times old in its center, and about 9 relaxation times at the half-mass radius. Therefore the cluster should be isotropic in the central part at least.
In this case, the fits have similar values of  $\chi^2$ and obtain similar results  for all values of $\beta_{out}$ at a given surface brightness profile. Our combined profile finds values for the black-hole mass  in the range of (1.9$-$2.3)$\cdot$10$^3$\,M$_\odot$ and M/L=1.07$-$1.09 at all values of $\beta$. 	Figure~\ref{fig:sphcont} is a contour plot of  $\chi^2$ as a function of $\beta$ and M$_\bullet$ from the combined surface brightness profile. For these models we use a constant and global M/L = 1.1, and do not fit the best value of M/L, since we are interested in showing the influence of $\beta$ and M$_\bullet$ on the quality of the fit. If  M/L is fitted, it usually has a value around 1.1 anyway.   In the case of radially varying anisotropy the abscissa denotes the value of $\beta_{out}$, the anisotropy in the outermost part of the cluster. The value of  $\chi^2$ is labeled on the contours. For constant $\beta$ the area with the best models is smaller than for $\beta \propto r$, where fits over a large range of $\beta$ are of similar goodness. We  conclude that anisotropy in the outer part of the cluster has no big effect on the  resulting black-hole mass in Jeans models as expected, but the central value of anisotropy is crucial.

From the $\Delta \chi^2$ = 1 limit we obtain an estimate of the uncertainty of a single model. However, to estimate the error of M/L and M$_\bullet$ we also run Monte Carlo simulations. In 1,000 runs we vary the shape of the combined  surface brightness profile to estimate the effect on the outcome of the Jeans models. We change only the six innermost points of the profile, as they have the largest uncertainties and are crucial to determine the mass of a central black hole. This is done for $\beta$ in the range of $-$0.20 to 0.20. 
 From the 68\% confidence limit we obtain the uncertainties  $\delta$M/L$_V$ = 0.01 and  $\delta$M$_\bullet$ = 0.4\,$\cdot$\,10$^3$\,M$_\odot$. For a confidence limit of 95\% we find  $\delta$M/L$_V$ = 0.08 and  $\delta$M$_\bullet$ = 1.2\,$\cdot$\,10$^3$\,M$_\odot$. 
 By varying both, the surface brightness profile and the velocity-dispersion profile, we run another Monte Carlo simulation and find  higher uncertainties of   $\delta$M/L$_V$ = 0.04 and   $\delta$M$_\bullet$ = 1.0\,$\cdot$\,10$^3$\,M$_\odot$ for the 68\% confidence limit, and  $\delta$M/L$_V$ = 0.16 and   $\delta$M$_\bullet$ = 4.0\,$\cdot$\,10$^3$\,M$_\odot$
with 95\% confidence.
The difference of the uncertainties by varying both profiles or only the surface brightness profile give information on the  contribution from the different profiles. The uncertainty of M/L seems to come mostly from the uncertainty of the kinematic profile, and the influence on the black-hole mass is also dominated by the kinematic profile. Nevertheless, one must not underestimate the importance of the shape of the surface brightness profile. 
 The $\Delta \chi^2$ = 1 limit of a single model gives also uncertainties in the same order of magnitude as the 68\% confidence limits, though some models have a larger range of uncertainty than from Monte Carlo simulations. 
	\begin{figure*}
	\centering
	\includegraphics[width=17cm]{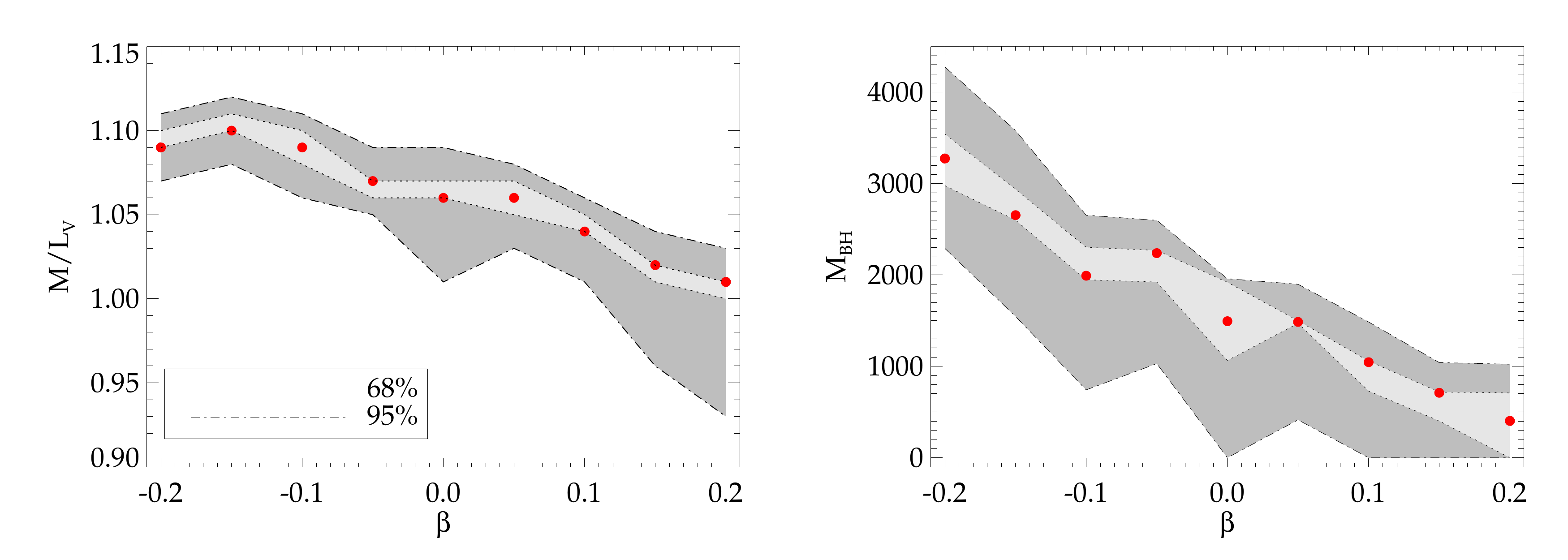}
			\caption{The result of Monte Carlo simulations on the surface brightness profile. After 1000 realizations for every value of constant $\beta$ the  dots denote the median values,  the   contours are the 68\% (dotted line) and 95\% (dash-dotted line)  confidence limits. The left panel displays the mass-to-light ratio M/L$_V$, the right panel is the black-hole mass.}
		\label{fig:sbpercSB}
		\end{figure*}
  Figure \ref{fig:sbpercSB} shows the median values of  these 1,000 realizations of the surface brightness profile as red  points and as a function of $\beta$, and the 68\% and 95\% confidence limits are the light grey and grey shaded contours. The uncertainty of M/L  is plotted in the left panel, M$_\bullet$ is plotted in the right panel. 
	
All our results come from Jeans models in which we neglect the PSF effects as described in  Appendix A of \cite{JAM}. To test the importance of this effects we run spherical Jeans models  and convolve the model velocity dispersion with the instrumental PSF before fitting to the observed data points. We use the combined surface brightness profile and constant values of anisotropy. With a seeing of 0.6\arcsec and a pixel scale of 0.52\,\arcsec/pixel the values for  black-hole mass, total cluster mass, and mass-to light ratio decrease by   less than 0.4\%. As the convolution is time consuming but has no big effect on the result, we disregard it in all further Jeans models.
	
\subsection{Spherical Jeans models with a varying M/L profile}	
\label{sec:sphml}
\begin{figure}
	\resizebox{\hsize}{!}{\includegraphics{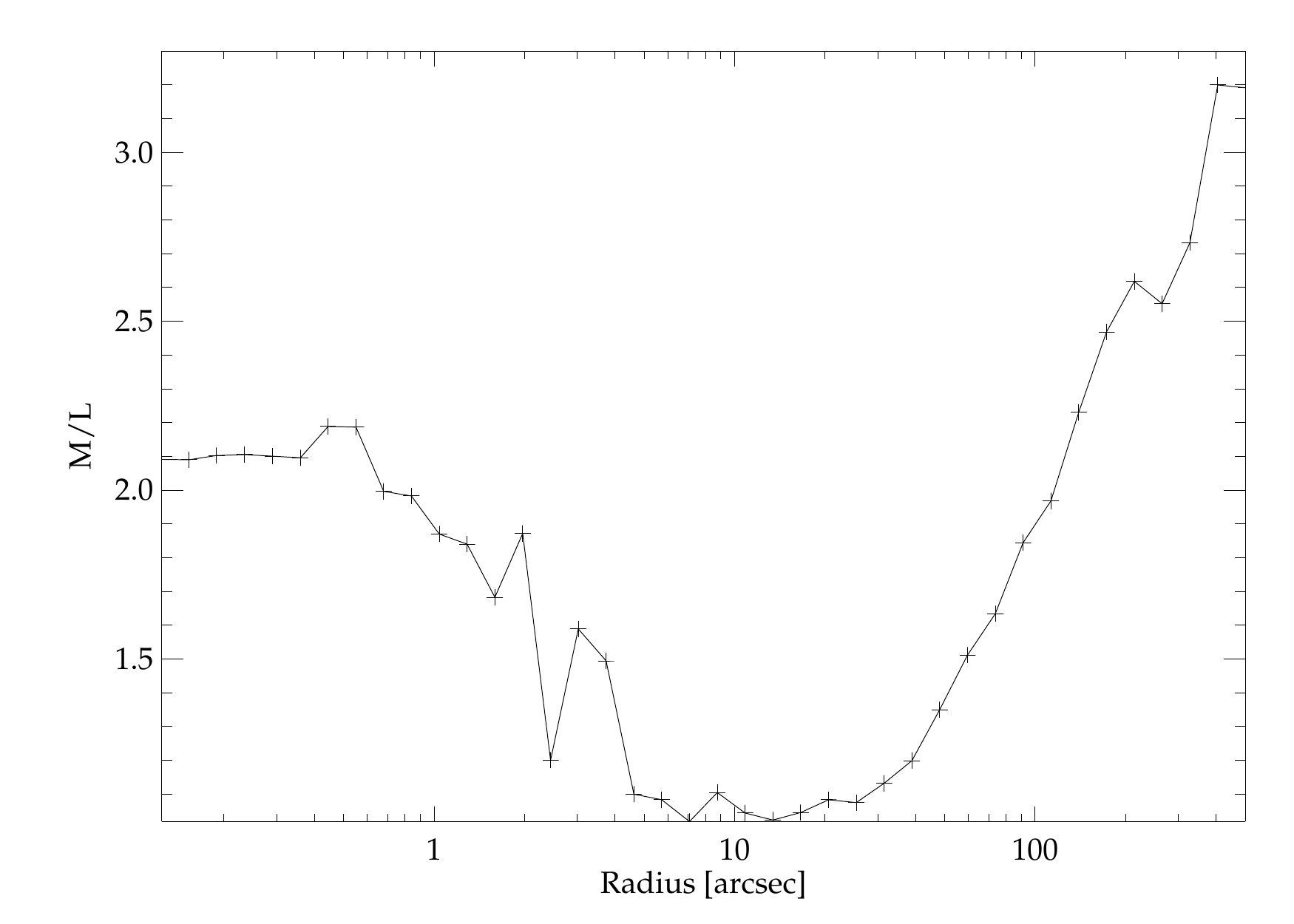} }
	\caption{Mass-to-light ratio as a function of radius, calculated from the best fitting $N$-body model.}
	\label{fig:mlnb}
	\end{figure}

$N$-body simulations of globular clusters usually result in mass-to-light ratios that vary with radius. This is  expected from mass segregation of the stars, as more massive stars, in particular dark remnants, wander to the center of the cluster, and low-mass stars  to the outskirts. We want to investigate the effect of a varying M/L profile on the Jeans models, and therefore calculate it from $N$-body simulations, and use it as   input to the Jeans models. The profile is displayed in Figure \ref{fig:mlnb}, it comes from the best fit of a grid of  $N$-body simulations, not including the effect of a tidal field (see Section \ref{sec:nbgrid}), to our profile from a combination of star counts and integrated light.	This profile is scaled to the velocity-dispersion profile in the Jeans models to obtain absolute values for M/L in the V-band. 
We use only the  combined profile, as it gives the best fit for constant M/L, and this profile was also used to determine the shape of the M/L profile.

With a constant value of $\beta$ = 0, the best-fit black-hole mass is (2.0\,$^{+1.3}_{-1.0}$)\,$\cdot$\,10$^3$\,M$_\odot$ with $\chi^2$ = 2.21 and the mean value of M/L is $\sim$\,1.5. A better fit is obtained with a constant anisotropy of $\beta$ = 0.15 with 
(0.9\,$^{+1.5}_{-0.9}$)\,$\cdot$\,10$^3$\,M$_\odot$, 	
M$_{tot}$ = (4.1\,$\pm$\,0.2)\,$\cdot$\,10$^5$\,M$_\odot$ and $\chi^2$ = 0.99.
 Except for the total cluster mass, the results are in agreement with the spherical Jeans models with constant M/L, they yield  similar masses for a central black hole. The higher cluster mass comes from the M/L profile, which  has higher values of M/L at small radii, where our surface brightness profile has the largest error bars. It also rises towards the outskirts, but our velocity-dispersion measurements extend only to 75\arcsec. Therefore the variable M/L profile adds mass to the  cluster without having a big effect on  the outcome of the Jeans models for M$_\bullet$.

\subsection{Axisymmetric Jeans models with constant M/L}

The axisymmetric model requires the 2D surface brightness profile, which we obtain with the \emph{MGE\_fit\_sectors IDL} package  written by  \cite{MGE}. We use the HST image and three 2MASS images in the J-band, which were combined to one image using the tool \emph{Montage}. The routine \emph{find$\_$galaxy.pro} measures the ellipticity $\epsilon$ = $1-b/a$ from a two dimensional image. One parameter of \emph{find$\_$galaxy.pro} is the fraction of image pixels that are used to measure the outcome. We use different fractions  (0.1, 0.15, 0.2, 0.25, and 0.3), and compute the mean values for ellipticity, which is 0.12\,$\pm$\,0.02 in the HST, 
 and 0.14\,$\pm$\,0.04 in the 2MASS image. 

Both values are in agreement with the value from \cite{white} of $\epsilon$ = 0.12.
The routine \emph{sectors\_photometry.pro} determines a 2D surface brightness profile from the HST and 2MASS images. The 2MASS photometry is matched  to the HST flux in the range from 2\arcsec\,to 100\arcsec, and scaled to the  HST flux. The HST measurements are further used for the central part of the cluster up to 60\arcsec\,, the 2MASS photometry for the outer part (13\arcsec $-$ 400\arcsec).
The photometry is then scaled to  the \cite{Trager} profile at  25\arcsec $-$ 158\arcsec. 
The routine \emph{MGE\_fit\_sectors.pro} fits a 2D MGE model to the measurements, and to every Gaussian of the fit there is an axial ratio 0 \mbox{$\leq$ $q _{k}$ $\leq$ 1} assigned. The total luminosity is L$_{tot}$=2.67\,$\cdot$\,10$^5$\,L$_{\odot}$. Varying the range where the photometry is scaled to \cite{Trager} introduces an  uncertainty  of  9\,$\cdot$\,10$^3$\,L$_{\odot}$.

Axisymmetric Jeans models have an additional input parameter, the inclination angle $i$, which is the angle between the rotation axis and the line-of-sight. An inclination angle $i$\,=\,90\degr\,means the cluster is seen edge-on, and $i$\,=\,0\degr\,corresponds to face-on. But in the face-on case we would probably see no indication of rotation, as we do in the outer part of the cluster, and the ellipticity $\epsilon$ would be zero. 
 We run axisymmetric Jeans models for $i$ =  45\degr, 67\degr, and 90\degr, and  constant $\beta$ in the interval of [-0.2, 0.3].  M/L is assumed to be constant and fitted in every model. 

\begin{table*}
\caption{The axisymmetric Jeans model best-fit values.}
 \label{tab:ax}
 \centering
\begin{tabular}{crccccc}
\noalign{\smallskip}
\hline\hline
\noalign{\smallskip}
$i$ &	$\beta$&	$M/L$&	$M_\bullet$&	$M_{tot}$&	$L_{tot}$&	$\chi^2$\\
$ \left[  \degr \right]  $& constant	& $ \left[  M_\odot/L_{V,\odot}  \right]  $ &	$ \left[  10^3\,M_\odot \right] $& 	$ \left[  10^5\,M_\odot \right]  $ &$\left[  10^5\,L_\odot \right]  $&\\
\noalign{\smallskip}
\hline
\noalign{\smallskip}

45 & $-0.10$&  $1.08 ^{+0.03}_{-0.05}$ & $1.5 ^{+1.6}_{-1.1}$&  $2.88 \pm 0.13$ & $2.67 \pm 0.09$&$1.16$ \\
&&&&&&\\
67 & $0.00$& $1.04 ^{+0.03}_{-0.05}$ & $1.9 ^{+1.4}_{-1.1}$&  $2.76 \pm 0.13$ &  $2.67 \pm 0.09$&$1.36$ \\
&&&&&&\\
90 & $0.00$&  $1.04 ^{+0.03}_{-0.05}$ & $1.9 ^{+1.4}_{-1.1}$&  $2.76 \pm 0.13$ &  $2.67 \pm 0.09$&$1.39$ \\
\noalign{\smallskip}
  \hline
\noalign{\smallskip}
\end{tabular}
\end{table*}

The  best fits at every value of $i$ are listed in Table \ref{tab:ax}. All errors are from the $\Delta \chi^2$=1 limit, the uncertainty in M$_{tot}$ comes  from the uncertainty in M/L and L$_{tot}$. At lower inclination we obtain better fits, and in contrast to spherical Jeans models, the best fit is at tangential anisotropy ($\beta$ = $-$0.1). At higher values of $i$ the isotropic case fits best. But all models obtain similar results for M/L, M$_\bullet$, and M$_{tot}$, which agree within their uncertainties, and are also consistent with the results from the spherical Jeans models.

\section{$N$-body simulations}
\label{sec:nb}
\subsection{Fitting the data to a grid of $N$-body simulations} 
\label{sec:nbgrid}
In addition to Jeans models we use a  grid of $N$-body models and fit the models to the data of NGC~5286. The simulations are  described in  \cite{mcnamara}. All  simulations use the code NBODY6 (\citealt{nbody6}) and start from \cite{king62} models with a \cite{kroupa}  initial mass  function between 0.1 and 100\,M$_\odot$. The stars evolve according to the stellar evolution model of  \cite{hurley}. The simulations run for 12 Gyr and we use 10 snapshots   with a 50\ Myr interval in between, starting at 11\,Gyr. We overlay the snapshots for comparison with the data. The cluster is then scaled up to NGC\ 5286 using the scaling as described by \cite{behrang}. 
To obtain smooth surface brightness and velocity-dispersion profiles from the simulations we use the infinite projection method of  \cite{infpro}, which averages the stellar positions over all possible rotations of the cluster relative to the observer.

Simulations cover a three-dimensional grid with different values of black-hole mass, concentration $c$ = $\log(r_t/t_c)$, and initial half mass radius R$_{Hi}$, and we search within this grid to obtain the model which best fits the observational data. The simulations are run without an external  tidal field, so cluster evolution is driven  by two-body relaxation and stellar evolution.  
For the fit to the $N$-body model we use our  combined surface brightness profile, as it also  results in the lowest $\chi^2$ with spherical Jeans models. The error bars in the outskirt of the profile are less than 1\%, which gives too much weight on the fit of the surface density. Therefore we use the approach of  \cite{Laughlin} to estimate the error  $\sigma_i$ of the \cite{Trager} profile from the relative weights of each data point ($\omega_i$) and a constant $\sigma_\mu$, with $\sigma_i$\,=\,$\sigma_\mu/\omega_i$. The weights were estimated by \cite{Trager}, and \cite{Laughlin} give the value of $\sigma_\mu$\,=\,0.128. We use these errors   for the combined surface brightness profile, starting at $\sim$\,3\arcsec. 
\begin{figure*}
	\centering
		\includegraphics[width=17cm]{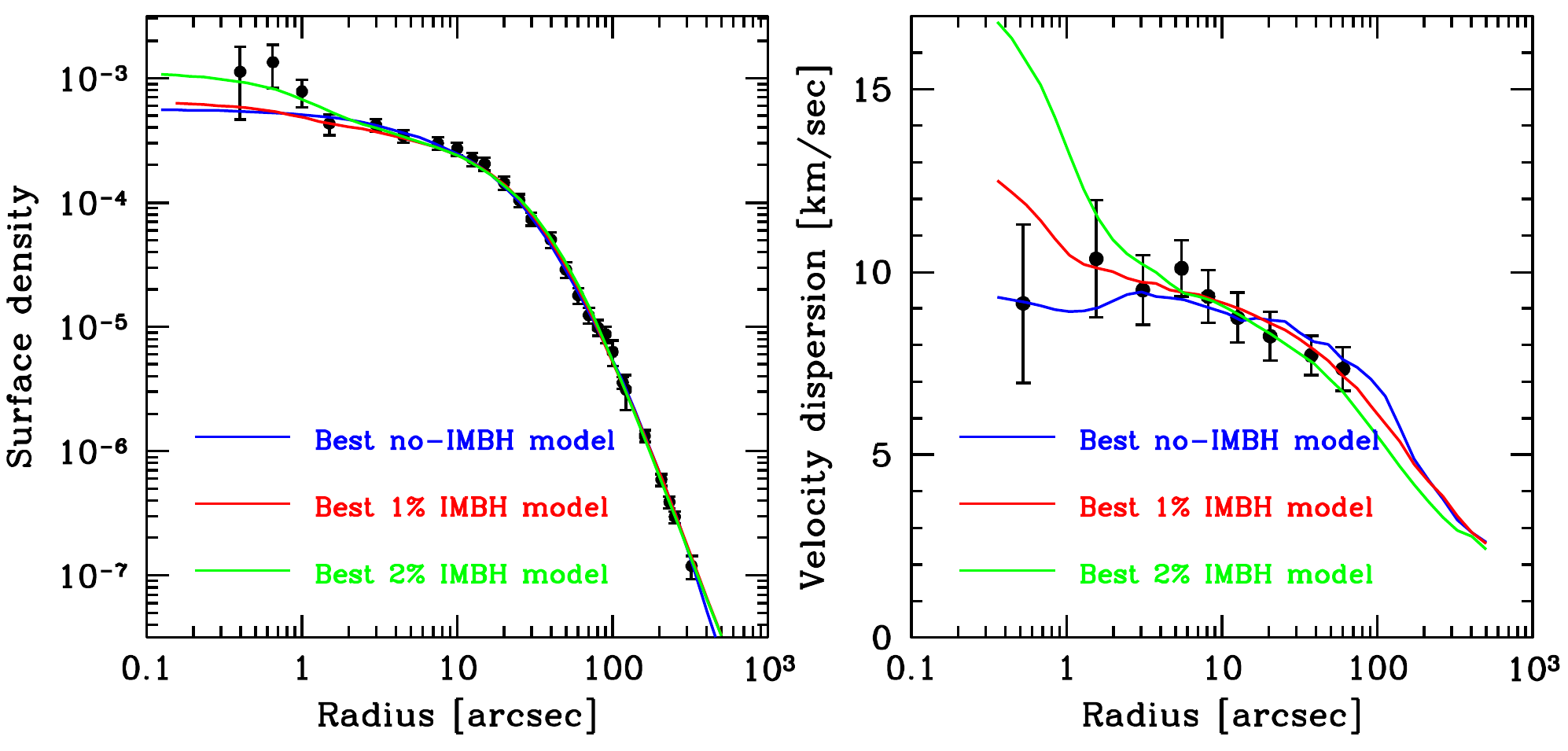}
	\caption{Best fitting $N$-body model without tidal field and with outer kinematic data. The surface density is plotted on the left panel, the velocity dispersion on the right panel. The different colors denote the best models for a black hole with 1\% of the cluster mass (red), 2\% of the cluster mass (green), and for no black hole (blue).  The data points are denoted as black symbols.}
	\label{fig:nbprof}
	\end{figure*}

The best fitting models for no IMBH, an IMBH with 1\%, and with 2\% of the total cluster mass are displayed in Figure \ref{fig:nbprof}. The surface density is shown in the left panel, and the velocity dispersion in the right panel. We convolved the model profiles to take the effects of seeing and pixel size into account. The change of the model profile is small except for  the center where R\textless\,0.5\arcsec, beyond our data points.
 All models have similar results for the surface brightness at radii R \textgreater\,5\arcsec\,and  differ only at the center. Regarding only the surface brightness profile, the 2\% IMBH model fits better than the 1\% IMBH or the  no-IMBH model. However, the velocity-dispersion profile of the 2\% IMBH model rises too much in the center of the cluster, and does not fit the innermost data point.  The 1\% model also rises towards the center, but not as much.  The best  overall fit  contains an IMBH with M$_\bullet$/M$_{tot}$ = 0.9\% and $\chi^2$\,=\,11.3 ($\chi^2_{red}$\,=\,0.39). This model gives a  total cluster mass of M$_{tot}$\,=\,(4.38$\,\pm\,$0.18)\,$\cdot$\,10$^5$\,M$_{\odot}$, therefore the IMBH has 
M$_\bullet$\,=\,(3.9$\,\pm\,$2.0)\,$\cdot$\,10$^3$\,M$_\odot$.
The best-fitting no-IMBH model has M$_{tot}$\,=\,4.48\,$\cdot$\,10$^5$\,M$_{\odot}$  with $\chi^2$\,=\,13.4. This case can be excluded at the 68\% confidence limit.
 
 \begin{figure*}
	\centering
	\includegraphics[width=17cm]{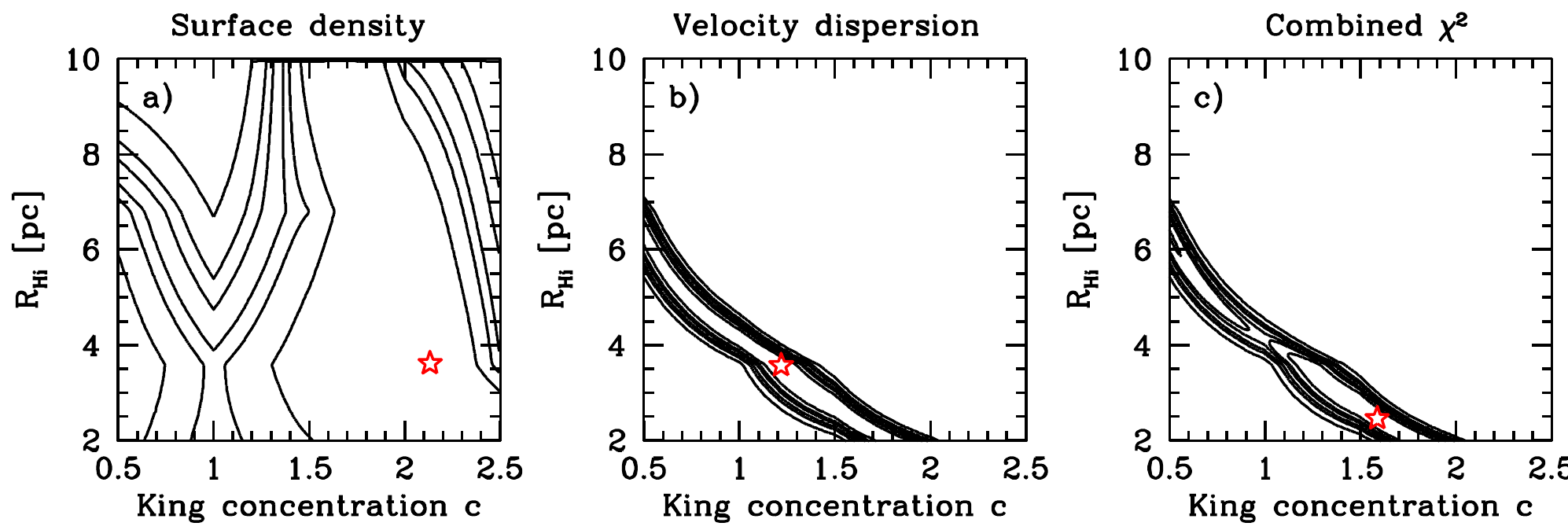} 
	\caption{$\chi^2$ contour lines as function of start parameters King concentration $c$ and initial half-mass radius R$_{Hi}$ for the surface density (a), velocity dispersion (b), and both profiles combined (c). The red star indicates the best fit.}
	\label{fig:nbchi}
	\end{figure*}

 Figure \ref{fig:nbchi}	 shows the $\chi^2$ contour lines as a function of the start parameters of the simulations for a 1\% IMBH model. The lowest $\chi^2$ for the surface density are at a region of models which start at a concentration of about $c=2$ (panel a). The value of the initial half-mass radius R$_{hi}$ is not much constrained. 
 For the velocity dispersion (panel b), best fits  are in a narrow band going from R$_{Hi}$\,=\,4.5\,pc and low concentration  to higher concentration models and R$_{Hi}$\,=\,2\,pc.  \cite{mcnamara} observe the same for NGC 6266 and argue that this is due to the dynamical evolution and the performed scaling of the cluster. However, the combined reduced $\chi^2$, considering the fit to both profiles (panel c) comes from a  region around $c$\,=\,1.6 and R$_{Hi}$\,=\,2.1\ pc and has the value $\chi^2_{red}$\,=\,0.39.

\subsection{Testing the spherical Jeans models with an $N$-body simulation}
\label{sec:testjeans}
In  order to test whether Jeans models give reliable results for the total cluster mass,  we use the $N$-body simulation which is closest to the best-fitting one and analyze it in more detail in this section.   This simulation started with an IMBH mass of 1\%, a half-mass radius is R$_{Hi}$\,=\,2.0\,pc, and an initial concentration of $c$\,=\,1.5. The simulation leads to a  cluster mass of 4.94\,$\cdot$\,10$^5$\,M$_{\odot}$. We compute the surface brightness and velocity-dispersion profiles for this simulation and perform a spherical Jeans analysis with these profiles. To compute the velocity dispersion  we make three perpendicular projections and compute a velocity-dispersion profile for each projection with the method of \cite{maxlik}. The three projections are averaged to a final velocity-dispersion profile.
We also vary the limiting magnitude of stars that contribute to the velocity-dispersion profile. Fainter stars tend to be at larger radii due to mass segregation, and therefore increase the velocity-dispersion profile there. Computing the profile with  brighter stars only underestimates the velocity dispersion at large radii, and the profile, which rises to the center, is too steep. Radial anisotropy also steepens the velocity-dispersion profile.

\begin{figure*}
	\centering
	\includegraphics[width=17cm]{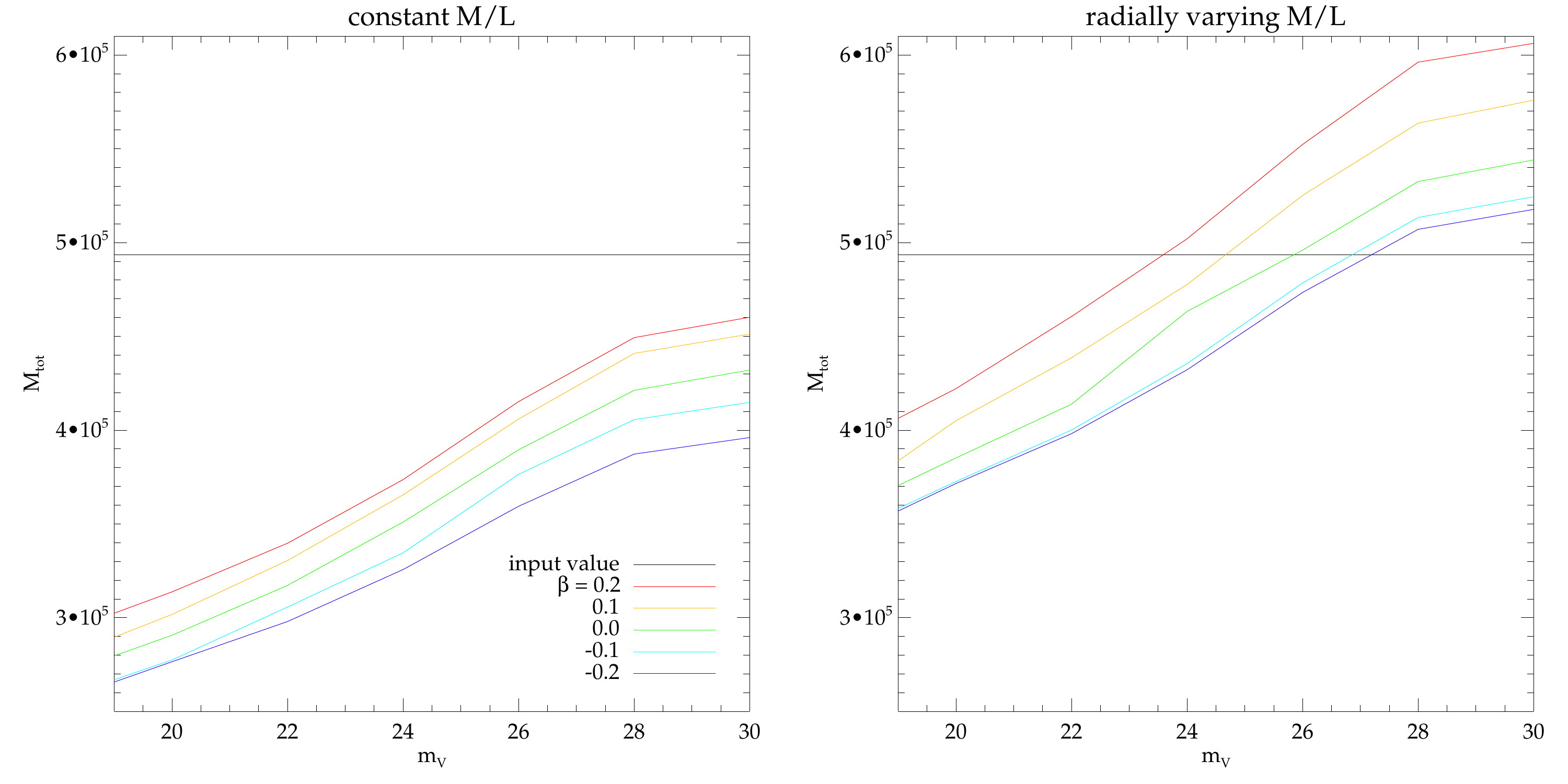}
		\caption{Total cluster mass M$_{tot}$ in M$_{\odot}$ from spherical Jeans models, performed on the profiles of an $N$-body simulation. For the surface brightness profile we use different limiting magnitudes m$_V$, thus changing the number of stars contributing to the profile. Left panel: constant M/L. Right panel: radially varying M/L profile. The solid black line is the input value of M$_{tot}$ in the simulation.}
	\label{fig:vlimtest}
	\end{figure*}

We run Jeans models with constant M/L and the radially varying M/L profile from the simulation, and different constant values of $\beta$. The limiting magnitude was chosen to be either m$_V$= 19, 20, 22, 24, 26, 28, or 30. The influence of varying the limiting magnitudes, the anisotropy parameter $\beta$, and the M/L profile on the total cluster mass is shown in Figure \ref{fig:vlimtest}.
The extracted velocity dispersion rises as a function of limiting magnitude. This results in a rise of total cluster mass as a function of limiting magnitude.
We run these tests with Jeans models of constant M/L as well as for radially varying M/L and find the same behavior for both. However, the recovered M$_{tot}$ is systematically lower for constant M/L, as constant M/L underestimates the cluster mass at large radii.  Also the value of $\beta$ influences the total cluster mass M$_{tot}$ found by the Jeans models. Tangential anisotropy results in higher M$_{tot}$ than radial anisotropy. The radially varying M/L profile at a magnitude cut of  m$_V$=26 and $\beta$=0 obtains M$_{tot}$\,=\,4.96\,$\cdot$\,10$^5$\,M$_{\odot}$, which is almost perfectly matching the input value of 4.94\,$\cdot$\,10$^5$\,M$_{\odot}$. As our spectral observations do not contain such faint stars, our measured velocity dispersion for NGC 5286 is probably also too low. This can be the reason why the Jeans models obtain a lower total cluster mass than the $N$-body simulation, even after taking the radially varying M/L into account.

\section{Conclusions}

\label{sec:end}
\subsection{Summary}
Using photometric and kinematic data, we modeled the globular cluster NGC\,5286 and found indications for the presence of a central black hole.
Our photometric data set consists of  an ACS/HST image, and the star catalog constructed by \cite{AndersonACS5} from ACS/HST observations. We used these to determine the photometric center with three different methods and confirmed the results of \cite{GoldCen}. Further, we produced a surface brightness profile from a combination of star counts and integrated light measurements. 
We used two kinematic data sets,  for the inner part of the cluster up to $\sim$\,25\arcsec\,we have ground based spectroscopic data from the  integral field unit ARGUS at the VLT. This data set was reduced and used to  produce a velocity map, and a velocity-dispersion profile. The second data set was obtained with the Rutgers Fabry Perot at the CTIO and contains velocity measurements of single stars. We used the outer part up to a radius of $\sim$\,75\arcsec\,to complement our velocity-dispersion profile. Examining both data sets for indication of rotation, we found rotation only in the region beyond 25\arcsec\,with a rotation velocity of (2.3 $\pm$ 1.5)\,km/s and a rotation axis consistent with the minor axis of the cluster.

With the information provided by the surface brightness profile we ran Jeans models and compared the results for the velocity dispersion to our profile. $\chi^2$ statistics were used to find the best fit. For the spherical models we used  our surface brightness  profile as well as the profiles from \cite{Trager}, and  \cite{NG06}. With different central black-hole mass and anisotropy  behavior the outcome of the model was scaled to our velocity-dispersion profile to determine the value of a constant mass-to-light ratio. Further, we used an M/L profile computed from $N$-body simulations as input. As a second approach we computed a  two-dimensional surface brightness profile  from the ACS/HST image and 2MASS images and ran axisymmetric Jeans models with different black-hole masses, anisotropy, and inclination. 
We also fit the surface brightness profile and the velocity-dispersion profile from a grid of $N$-body models without tidal field to our data.  
All our models require a central black hole for their best fit. Jeans models find a best fitting black-hole mass of about (1.5$\pm$1.0)\,$\cdot$\,10$^3$\,M$_\odot$, while N-body modes require a black hole of 0.9\% of the total cluster mass, corresponding to M$_\bullet$ = (3.9$\pm$2.0)\,$\cdot$\,10$^3$\,M$_\odot$. The errors are the 68\% confidence limits. The models also provide the total mass of the cluster, and the ratio of the black-hole mass to the total cluster mass is lower for Jeans models,  M$_\bullet$/M$_{tot}$ $\approx$ 0.5\% $-$ 0.7\%. $N$-body simulations obtain a total cluster mass M$_{tot}$ =  (4.38$\,\pm\,$0.18)\,$\cdot$\,10$^5$\,M$_{\odot}$. The total cluster mass of spherical Jeans models is up to 34\% lower, depending on the value of M/L. For constant M/L we obtain a value of 1.07\,M/L$_V$ and  M$_{tot}$ =  (2.89$\,\pm\,$0.10)\,$\cdot$\,10$^5$\,M$_{\odot}$, but with the varying  M/L profile from $N$-body simulations the cluster mass is higher (M$_{tot}$= (4.1\,$\pm$\,0.2)\,$\cdot$\,10$^5$\,M$_\odot$).

\subsection{Discussion}
The value of  M/L from our Jeans models with constant M/L is about 1.07\,M$_\odot$/L$_{V,\odot}$. Both spherical and axisymmetric Jeans models find similar values. The surface brightness profiles from \cite{NG06} and \cite{Trager}  result in higher values of 1.11\,M$_\odot$/L$_{V,\odot}$ and 1.22\,M$_\odot$/L$_{V,\odot}$, respectively. This is probably due to the lower value of  total luminosity with these profiles (see Table \ref{tab:sphe}).    The integrated V-band  magnitude  of \cite{harris} is 7.34, corresponding to a total luminosity of 2.7\,$\cdot$\,10$^5$\,L$_\odot$.  This value is in good agreement with the total luminosity from our profiles. 

\cite{maxlik}  found a  value of M/L\,=\,2.1\,M$_\odot$/L$_{V,\odot}$ from integrated light measurements and isotropic King models.  \cite{Laughlin} used the \cite{Trager} profile and data from \cite{harris} to compute M/L ratios. They used a population-synthesis model (M/L = (1.87\,$\pm$\,0.16)\,M$_\odot$/L$_{V,\odot}$) and dynamical models to fit M/L (King fit: 0.99 $^{+0.49}_{-0.39}$, Wilson fit: 0.94 $^{+0.47}_{-0.37}$, and Power-law fit: 0.91 $^{+0.47}_{-0.35}$ in solar units, respectively), and obtained overall lower values than \cite{maxlik}, in better agreement with our  Jeans model with constant M/L of $\sim$1.07\,M$_\odot$/L$_{V,\odot}$.  \cite{maraston} also calculated M/L ratios from evolutionary population synthesis models. The results of the  Sloan Digital Sky Survey (SDSS) r-band filter at 622\,nm can be compared to our values of M/L$_{V}$ in the V-band at 606\,nm. With the \cite{kroupa} initial mass  function, a metallicity of $\lbrack $Z/H$\rbrack$ = $-1.35$ and an age of 11$-$12\,Gyr, their model gives   higher values of M/L=1.87$-$2.02. Our M/L profile from $N$-body simulations has a global value of M/L$_V$ of about (1.75\,$\pm$\,0.06)\,M$_\odot$/L$_{V,\odot}$ after scaling to the velocity-dispersion profile, which is in rough agreement with the M/L ratio of the population synthesis models. 

The differences in the total cluster mass between our  models can be explained by the influence of the M/L profile. With the M/L profile from $N$-body simulations  as input to our spherical Jeans models in Section \ref{sec:sphml} we obtain a higher cluster mass of M$_{tot}$\,=\,(4.1\,$\pm$\,0.2)\,$\cdot$\,10$^5$\,M$_\odot$,  in better agreement with the value found from the  $N$-body simulations of M$_{tot}$ =  (4.38$\,\pm\,$0.18)\,$\cdot$\,10$^5$\,M$_{\odot}$. Also the Jeans models that we ran on the profiles of an $N$-body simulations in Section \ref{sec:testjeans} confirm the influence of a radially varying M/L profile on the total cluster mass.
In the range of 2\arcsec \textless \,r \textless \,80\arcsec\, the M/L profile is lower than average. The extra mass, compared to a constant M/L, is mostly at larger radii. Since we have no velocity measurements for r \textgreater 75\arcsec, the additional mass at large radii does not significantly influence our velocity-dispersion measurements. A rise in M/L towards the outer cluster parts could also explain why the dynamical M/L values of \cite{Laughlin} are smaller than the population synthesis ones. This stresses the importance of obtaining velocity-dispersion measurements in the outer cluster parts when trying to determine the total cluster mass. Furthermore, the velocity-dispersion measurements are based on bright stars. Those stars are more massive than average, but have lower velocities. This underestimates the velocity dispersion and the total cluster mass, as we showed in Section \ref{sec:testjeans}.

For the determination of a central black-hole mass, the value  of the anisotropy $\beta$ is important. Anisotropy in the outer part of the cluster does not change the result much, but a radial anisotropy in the center can easily be misinterpreted as a central black hole, whereas unknown tangential anisotropy results in an underestimation  of the black-hole mass.  Assuming radial anisotropy of  $\beta$\,=\,0.30 in the center, a central black hole is not necessary to explain the velocity-dispersion profile of NGC\,5286.  To measure the amount of anisotropy, proper motion measurements are needed in addition to radial velocity dispersion. However, proper motion measurements require a long time baseline and  high accuracy. Crowding in the center causes further problems. Therefore the only reliable method to constrain the anisotropy  are $N$-body simulations.   \cite{Nora} ran a simulation of a highly  anisotropic cluster and showed that the central part of a cluster becomes isotropic  after a few relaxation times. As NGC 5286 is already 9 relaxation times old at its half-mass radius, we expect not much anisotropy ($\left| \beta \right|$ \textless\, 0.2). Especially in its central part, where the relaxation is even faster, the cluster should be isotropic.
Nevertheless our Jeans models obtain better fits for radial anisotropy than for isotropy. Mass segregation could be one reason to explain this: The low-mass stars, which are mostly in the outer part of the cluster, are too faint for detection, and the velocity-dispersion profile at larger radii can therefore be underestimated compared to inner radii. Our velocity-dispersion profile may be too steep, and the Jeans models misinterpret this as radial anisotropy.

The Jeans models  find smaller values for the mass of the black hole  than the $N$-body models, but  both indicate a black hole with less than 1\% of the total cluster mass. This upper limit is higher than the  black hole mass fraction usually measured for galaxies.    \cite{nadine} found that the fraction of a central massive dark object in nearby galaxies is around 0.14\%\,$\pm$\,0.04\% of the bulge mass of the host system.

We have 1-$\sigma$ detections for an IMBH, so the no-black-hole case cannot be excluded. A central black hole should not be more massive than 6.0\,$\cdot$\,10$^3$\,M$_{\odot}$, as for $N$-body simulations this is where $\Delta\chi^2$ = 5. Jeans models find an upper limit of M$_\bullet$ = 5.5\,$\cdot$\,10$^3$\,M$_{\odot}$ with 95\% confidence level from Monte Carlo simulations. But independent of the method we use, all our methods prefer the case with an IMBH over the no-black-hole case. The result from $N$-body simulations is M$_{\bullet}$=(3.9 $\pm$ 2.0)\,$\cdot$\,10$^3$\,M$_{\odot}$, from Jeans models the black-hole mass is about 50\% lower.

Our modeling shows that the derived uncertainties for the kinematic profile are too high to precisely  constrain the black-hole mass and anisotropy. In the innermost bin the uncertainty is more than 2.2 km/s, which is 24\% of the measured value. However, these high error bars are realistic, as the determination of the velocity dispersion currently shows some ambiguity. We used two different methods and varied the parameters of the \emph{pPXF} method. Thus we  found scatter of up to 3\,km/s.  Further investigation of this inconsistency is therefore needed to obtain better constrained results. Alternatively, observations with higher spectral resolution or of individual stars could also  clear up this issue.

NGC 5286 is an interesting case for dynamical modeling and could host an intermediate-mass black hole. Our research group has  IFU data  for six more  globular clusters in the southern hemisphere \citep{6cluster}.  These data are used to constrain possible IMBHs inside these clusters. 
This information can be used to decide which other clusters should be examined, also in the northern hemisphere. Located between galaxies and globular clusters, at $\sigma_{e}$\,$\sim$\,15~-~70\,km/s, are systems like dwarf galaxies, which could also contain  intermediate-mass black holes.
Assessing  the fraction of globular clusters and stellar systems with intermediate-mass black hole, the  masses of the black holes and their environments, will help us to understand not only  the dynamical processes in these stellar systems, but maybe also  the formation of supermassive black holes.

 \begin{acknowledgements}
We thank the anonymous referee for her/his comments and suggestions, which really helped to improve our manuscript.  This research was supported by the DFG cluster of excellence Origin and Structure of the Universe (www.universe-cluster.de). 
 H.B. acknowledges support from the Australian Research Council through Future Fellowship grant FT0991052.
This research made use of Montage, funded by the National Aeronautics and Space Administration's Earth Science Technology Office, Computation Technologies Project, under Cooperative Agreement Number NCC5-626 between NASA and the California Institute of Technology. Montage is maintained by the NASA/IPAC Infrared Science Archive. 
 \end{acknowledgements}
 \bibliographystyle{aa}		

\bibliography{bibs}		

\end{document}